\documentclass[12pt]{article}
\usepackage{amsfonts}
\usepackage{amsmath, amsthm}
\usepackage{epsfig}
\usepackage{algorithm}
\usepackage{algorithmic}
\usepackage{epic}
\usepackage{threeparttable}

\usepackage[a4paper]{geometry}
\usepackage{enumerate}
\usepackage{amsmath,verbatim,color,amssymb,epsfig}
\usepackage{bm}
\usepackage{ragged2e}
\usepackage{amsfonts}
\usepackage{epsfig}
\usepackage{changepage}
\usepackage{multirow}
\usepackage{graphicx}
\usepackage{array}
\usepackage[table]{xcolor}
\definecolor{Gray}{gray}{.80}
\usepackage{lscape}

\usepackage[round,semicolon,authoryear]{natbib}
\usepackage[normalem]{ulem}
\usepackage[colorlinks=true,urlcolor=blue,citecolor=purple,linkcolor=blue,bookmarks=true]{hyperref}
\usepackage{setspace}

\usepackage{booktabs}  
\usepackage{cleveref} 
\usepackage{caption}  
\usepackage{tabularx}   
\usepackage{adjustbox} 
\usepackage{longtable}  

\begin{document}
\doublespacing   
\baselineskip=12pt
\begin{center}
{\Large \bf A seamless dose-optimization design for monotherapy and combination therapy}
\end{center}
\vspace{1mm}
\begin{center}
{\bf Kentaro Takeda$^{1*}$, Masahiro Kojima$^{2}$}\\
\vspace{0.5cm}
\noindent $^{1}$Astellas Pharma Global Development Inc., Northbrook, IL, USA.\\
\vspace{1mm}
$^{2}$Chuo University, Bunkyo-ku, Tokyo, Japan.\\
\vspace{1mm}
$^*$Author for correspondence: kentaro.takeda@astellas.com
\end{center}
\vspace{2mm}

\baselineskip=24pt

\noindent \emph{\textbf{Abstract}}: The emergence of molecular-targeted agents and immune-oncology therapies has fundamentally transformed oncology drug development, necessitating evolution beyond traditional dose-finding approaches designed for cytotoxic agents. While conventional agents exhibit predictable monotonic dose-response relationships, novel anticancer agents often demonstrate plateau-effect patterns where higher doses may compromise therapeutic benefit, requiring identification of optimal biological doses that balance efficacy and tolerability. The FDA's Project Optimus initiative emphasizes comprehensive dose optimization through parallel randomized cohorts and patient backfilling to better understand pharmacological profiles across multiple dose levels. Contemporary drug development increasingly prioritizes combination therapy alongside monotherapy evaluation, yet existing designs typically assume equivalent roles for both agents, diverging from clinical practice where novel agents combine with established treatments having limited dose options. This paper proposes a seamless dose-optimization design that adaptively evaluates both monotherapy and combination therapy based on efficacy and toxicity outcomes through adaptive subtrials with patient backfilling capabilities. The model-assisted framework employs predetermined Bayesian optimal boundaries, eliminating real-time model fitting while accommodating evaluation of both monotherapy and combination therapy and enabling sequential enrollment with strategic backfilling. Simulation studies demonstrate robust performance across diverse dose-response patterns relevant to contemporary oncology. The design addresses critical gaps between methodological assumptions and clinical reality, offering a practical approach that integrates monotherapy and combination therapy evaluation with efficacy-toxicity-based backfilling. This framework ensures straightforward implementation while maintaining flexibility for complex dose-response relationships characteristic of novel anticancer agents, providing significant advantages for clinical implementation in the era of targeted and immune-oncology therapeutics.

\vspace{0.5cm}

\noindent \emph{\textbf{keywords}}: dose-optimization; monotherapy; combination therapy; model-assisted design; backfill.


\section{Introduction}\label{sec_intro}
Oncology drug development has undergone a fundamental transformation with the emergence of molecular-targeted agents and immune-oncology therapies. This paradigm shift has necessitated a corresponding evolution in dose-finding methodologies. Traditional cytotoxic agents follow predictable patterns, making the maximum tolerated dose (MTD) a reasonable target. In contrast, novel anticancer agents exhibit complex and often unpredictable dose-response relationships that challenge conventional dose-finding approaches. The primary objective for these agents is to identify an optimal biological dose (OBD)—a tolerable dose that achieves adequate therapeutic efficacy despite uncertain dose-toxicity and dose-efficacy relationships.
The dose-efficacy profiles of molecular-targeted and immune-oncology agents frequently deviate from traditional assumptions due to distinct mechanisms of action. Molecular-targeted agents often exhibit plateau effects because their efficacy depends on binding to specific molecular targets; once these targets are saturated at intermediate doses, additional drug exposure provides no incremental therapeutic benefit while potentially increasing off-target toxicities~\citep{Druker2002-ms}. Immune-oncology agents can demonstrate even more complex, sometimes bell-shaped, dose-efficacy relationships due to the intricate dynamics of immune activation. At optimal doses, these agents effectively stimulate anti-tumor immune responses, but higher doses may paradoxically compromise efficacy through mechanisms such as T-cell exhaustion, induction of regulatory T-cells, or excessive inflammatory responses that suppress immune function~\citep{Postel-Vinay2009-lu}. 
These complex relationships have prompted the development of sophisticated dose-finding methodologies that simultaneously consider both efficacy and toxicity endpoints. Numerous model-based designs~\citep{Thall2004-ad, Yuan2017-ng} and model-assisted approaches~\citep{Takeda2018-rs,Lin2020-vf,Yuan2022-fm} have emerged to address these challenges in early-phase oncology trials.

The FDA Oncology Center of Excellence launched the Project Optimus to reform the paradigm of dose optimization and dose selection in the development of oncology drugs and issued guidance 'Optimizing the Dosage of Human Prescription Drugs and Biological Products for the Treatment of Oncologic Diseases'~\citep{US-Food-and-Drug-Administration2024-dr}. The project aims to ensure that oncology drug doses are optimized to maximize efficacy, safety, and tolerability and to understand the pharmacokinetics, pharmacodynamics, toxicity, and efficacy at each dose level. 
The FDA Project Optimus recommends that dose-finding trials include randomized parallel dose-response cohorts of multiple doses to generate these additional data at promising dose levels. Therefore, dose-finding trial designs including parallel randomized dose-response cohorts of multiple doses~\citep{Zhou2019-nh,Guo2023-ha,Takeda2023-ka} and backfilling patients at lower doses~\citep{Dehbi2021-um,Barnett2023-dm,Liu2024-vw,Zhao2024-ng,Pin2024-hi,Takeda2025-do,Zhao2025-hp,Kojima2025-vd,Chen2025-up} have recently been proposed. 

Contemporary oncology drug development increasingly prioritizes combination therapy alongside monotherapy evaluation. \citet{Kelly2024-oe} have introduced that the emergence of anticancer agents with distinct mechanisms of action has expanded therapeutic options and accelerated the development of combination therapies (Combo) aimed at achieving synergistic effects and overcoming drug resistance. In line with this trend, early-phase oncology trials increasingly evaluate Combo therapy alongside monotherapy (Mono). 
\citet{Coleman2026-cn} have discussed that the development of novel rational combination strategies is essential to overcome compensatory escape mechanisms and intrinsic and acquired resistance mechanisms, as well as other challenges, such as intratumoral heterogeneity and clonal evolution. 
The rationale for combining novel agents with established standard-of-care treatments has led to the development of specialized dose-finding designs for identifying optimal biological dose combinations (OBDCs). Existing approaches include both model-based~\citep{Cai2014-kj, Guo2015-wi, Lyu2019-ao} and model-assisted designs~\citep{Kakizume2024-sr, Lu2025-fb} that incorporate dual efficacy and toxicity endpoints. However, current combination therapy designs typically assume equivalent roles for both agents and similar dose level structures. This assumption diverges from real-world clinical practice, where investigators commonly conduct dose-finding trials for novel anticancer agents as monotherapy before exploring combinations with standard agents that have limited, well-established dose options.

To address this gap between methodological assumptions and clinical reality, this paper proposes a seamless dose-optimization design for monotherapy and combination therapy based on efficacy and toxicity outcomes. Our approach incorporates adaptive subtrials for monotherapy and combination therapy, enabling patient backfilling at lower doses while maintaining focus on dose escalation to explore higher dose levels. The model-assisted nature of the proposed design ensures predetermined decision rules that are both straightforward to implement and practical for clinical use, eliminating the need for real-time model fitting while accommodating the complexities of evaluation of both monotherapy and combination therapy and efficacy-toxicity-based backfilling. The design further enables sequential patient enrollment to accelerate trial conduct through strategic backfilling, offering significant practical advantages for clinical implementation.
Comprehensive simulation studies are presented to compare the operating characteristics of the proposed design with established dose-finding approaches across diverse realistic clinical scenarios. These evaluations demonstrate the design's performance across various dose-response patterns and clinical contexts relevant to contemporary oncology drug development.

The remainder of this paper is structured as follows. Section \ref{methodology} presents the proposed design framework, including the Bayesian optimal boundaries, the backfilling methodology based on cumulative and pending efficacy and toxicity outcomes, the subtrial-specific dose-finding algorithms, and the integrated dose-optimization process. Section \ref{simulation} details the simulation studies evaluating the proposed design's performance relative to alternative designs across multiple realistic scenarios. Section \ref{discussion} concludes with a summary of our findings and their implications for future oncology dose-finding practice.

\section{Methodology}\label{methodology}
\subsection{Motivating Example}\label{motivation} 
Our research has been motivated by a real phase I/II study evaluating a novel compound in combination with erlotinib for patients with EGFR activating mutation-positive (EGFRm+) advanced NSCLC who have developed resistance to EGFR tyrosine kinase inhibitor therapy. The study aims to identify the MTD and OBD for the novel compound as monotherapy, as well as the maximum tolerated dose combination (MTDC) and OBDC when combined with erlotinib. The novel compound is evaluated at five dose levels, while erlotinib is administered at two established doses (100 mg and 150 mg). The study follows a sequential approach: first establishing the MTD and OBD for the novel compound monotherapy, then determining the MTDC and OBDC for combinations with erlotinib at 100 mg, and finally evaluating combinations with erlotinib at 150 mg. Each dose-finding cohort begins at the lowest dose of the novel compound and proceeds independently. This conventional design presents several limitations: the restricted number of dose combinations precludes the use of existing complex dose-finding methods, the separate determination of MTD(C) and OBD(C) across dose-finding cohorts may result in information loss, and the requirement to start at the lowest dose for each dose-finding cohort leads to unnecessary patient exposure to potentially subtherapeutic doses. Such inefficiencies are commonly encountered in clinical practice, and our proposed approach aims to address these systematic challenges.

\begin{figure}[tbp]\centering
\includegraphics[width=1.10\linewidth]{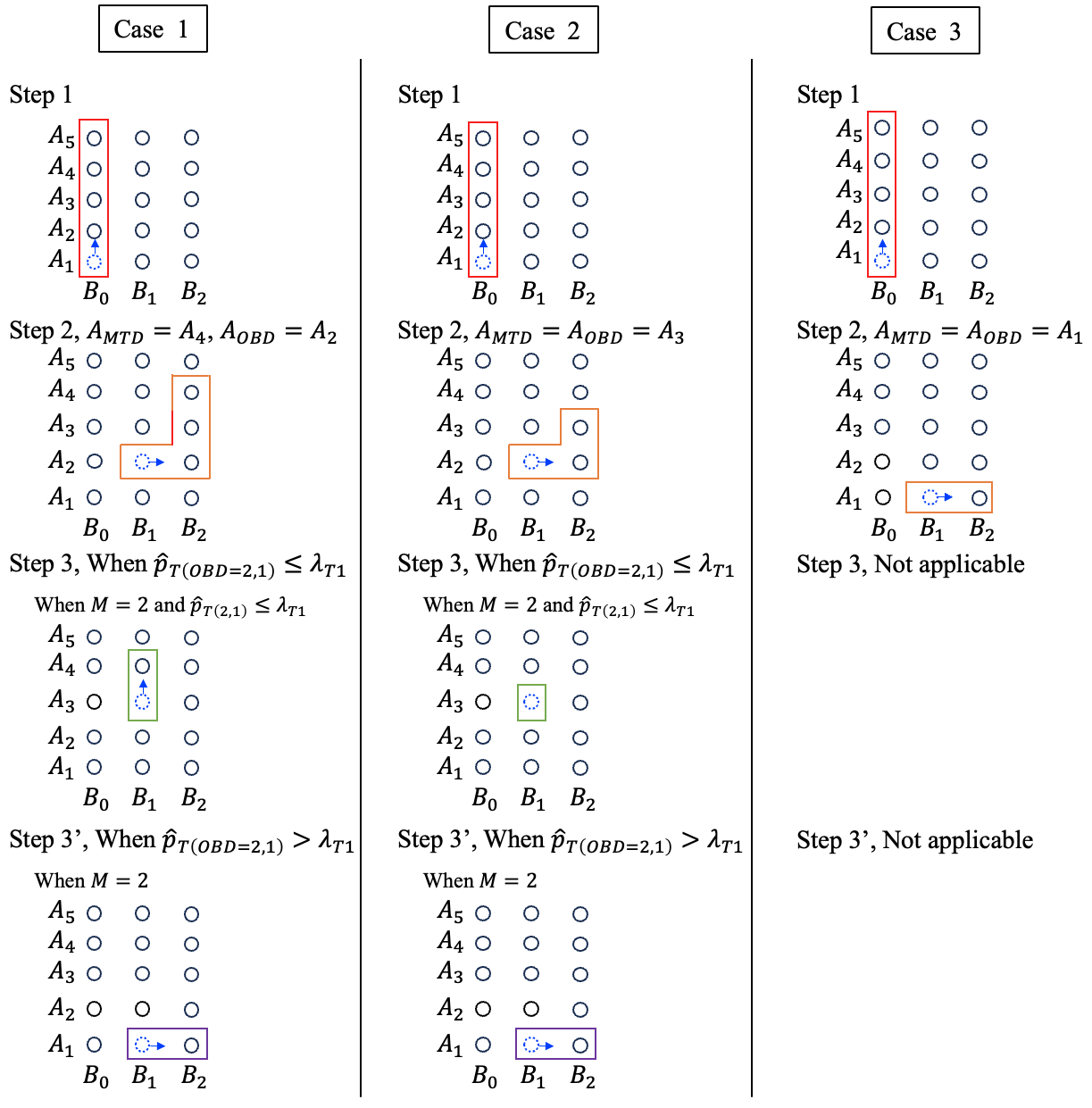}
  \caption{Examples of the dose adjustment procedure. $A_{l}$ and $B_{m}$ denote dose levels of agent A and agent B, respectively. Each subtrial is enclosed by colored lines. $\hat{p}_{T(l,m)}$ represents the estimated toxicity probability at dose combination $(A_{l}, B_{m})$, and $\lambda_{T1}$ is the lower Bayesian optimal boundary for toxicity.}
  \label{fig:dose_adjustment}
\end{figure}

Seamlessly optimizing doses for both monotherapy and combination therapy presents substantial challenges, as it requires navigating both one-dimensional monotherapy dose space and two-dimensional combination dose space simultaneously.
To address this complexity, we propose treating monotherapy doses and combination therapy dose combinations as a single dose matrix, enabling the evaluation of both treatment modalities within a unified framework. 
This dose matrix is systematically divided into subtrials to facilitate dose-finding for both monotherapy and combination therapy. 
We employ an adaptive partitioning strategy that ensures each subtrial maintains monotonic dose-toxicity relationships across dose levels. Information from completed subtrials serves solely to define subsequent subtrials and establish appropriate starting doses. Consequently, standard dose-finding algorithms can be applied within each subtrial to determine dose escalation decisions. 
The procedures are summarized in Figure~\ref{fig:dose_adjustment} and detailed in subsequent sections.
This approach streamlines implementation by avoiding the complex two-dimensional modeling and decision-making processes typically required by existing methodologies.

\subsection{Bayesian optimal boundaries}\label{boundaries} 
Suppose that the current cohort is treated at the dose level $j\:(j=1,...,J)$, where $n_{j}$ denotes the number of patients treated and $y_j$ represents the binary outcome at dose level $j$. 

For toxicity, referring to \cite{Liu2015-pc}, let $p_{T1} < p_{T2} < \ldots < p_{TJ}$ represent the true toxicity probabilities at dose levels $1, 2,...,J$, where the subscript $T$ denotes toxicity. Let $\phi_{T0}$ denote the target toxicity probability specified by the investigator, $\phi_{T1}$ the highest toxicity probability that is deemed subtherapeutic (warranting dose-escalation), and $\phi_{T2}$ the lowest toxicity probability considered overly toxic (requiring dose de-escalation).
Furthermore, let $\lambda_{T1}$ and $\lambda_{T2}$ denote the lower and upper cutoffs, satisfying $\phi_{T1}<\lambda_{T1}<\phi_{T0}<\lambda_{T2}<\phi_{T2}$. 
Let $\hat{p}_{Tj}=y_{Tj}/n_{j}$ denote the observed toxicity probability at dose level $j$, where $y_{Tj}$ denotes the toxicity outcome at dose level $j$.
The following dose allocation rules are assumed for toxicity: 
\begin{itemize}
\item If $\hat{p}_{Tj}\le\lambda_{T1}$, escalate the dose level to $j+1$;
\item If $\hat{p}_{Tj}>\lambda_{T2}$, de-escalate the dose level to $j-1$;
\item If $\lambda_{T1}<\hat{p}_{Tj}\le\lambda_{T2}$, stay at the current dose $j$.
\end{itemize}
By minimizing the probability of incorrect marginal decisions for toxicity, the optimal values of $\lambda_{T1}$ and $\lambda_{T2}$ can be determined as follows:
$$
\lambda_{T1}=\textup{log}\Big(\frac{1-\phi_{T1}}{1-\phi_{T0}}\Big)\Big/\textup{log}\Big(\frac{\phi_{T0}(1-\phi_{T1})}{(1-\phi_{T0})\phi_{T1}}\Big),
$$
$$
\lambda_{T2}=\textup{log}\Big(\frac{1-\phi_{T0}}{1-\phi_{T2}}\Big)\Big/\textup{log}\Big(\frac{\phi_{T2}(1-\phi_{T0})}{(1-\phi_{T2})\phi_{T0}}\Big).
$$

For efficacy, similar to \cite{Lin2017-na}, two simple hypotheses are considered to derive the optimal value for the efficacy probability. 
Let $p_{Ej}$ be the true efficacy probability at dose level $j$, where the subscript $E$ denotes efficacy, and let $\phi_{E0}$ and $\phi_{E1}$ denote the required efficacy probability specified by the investigator and an inefficacious efficacy probability, respectively. 
Let $\lambda_{E}$ be the efficacy cutoff, satisfying $\phi_{E1}<\lambda_{E}<\phi_{E0}$. 
Let $\hat{p}_{Ej}=y_{Ej}/n_{j}$ denote the observed efficacy probability, where $y_{Ej}$ denotes the efficacy outcome at dose level $j$.
For a tolerable dose level, the correct decisions from an efficacy perspective are described as follows. 
\begin{itemize}
\item If $\hat{p}_{Ej}\ge\lambda_{E}$, stay at the current dose level $j$; 
\item If $\hat{p}_{Ej}<\lambda_{E}$, explore other possible dose levels. 
\end{itemize}
By minimizing the probability of incorrect decisions about efficacy, the optimal value of $\lambda_{E}$ is determined as follows: 
$$
\lambda_{E}=\textup{log}\Big(\frac{1-\phi_{E1}}{1-\phi_{E0}}\Big)\Big/\textup{log}\Big(\frac{\phi_{E0}(1-\phi_{E1})}{(1-\phi_{E0})\phi_{E1}}\Big).
$$

The details of the Bayesian optimal boundaries are summarized in Supplementary \ref{sup_boundaries}.

\subsection{Likelihood with pending outcomes}\label{likelihood} 
In Section \ref{boundaries}, the observed probabilities of toxicity and efficacy, $\hat{p}_{Tj}$ and $\hat{p}_{Ej}$, based on completely observed data, are considered to estimate the Bayesian optimal boundaries. 
When a backfill approach is considered, the dose-escalation cohort exploring a higher dose usually enrolls patients first, followed by the backfill cohort. 
Therefore, $\hat{p}_{Tj}$ and $\hat{p}_{Ej}$ may not be available when the dose-escalation cohort has sufficient information and the next dose assignment is considered because the backfill cohort has not completed the toxicity and efficacy evaluation periods. 
To address this issue, the likelihood with pending data \citep{Lin2020-xx,Takeda2020-yc} is used.

The observed toxicity data are Bernoulli variables $\tilde{y}_{Ti}, i=1,...,n_{j}$, which indicate that the patient has experienced toxicity ($\tilde{y}_{Ti}=1$), or has not yet experienced toxicity ($\tilde{y}_{Ti}=0$), at the decision times for dose-escalation or de-escalation, say $k$. 
Here, $\tilde{y}_{Ti}=1$ implies $y_{Ti}=1$, but $\tilde{y}_{Ti}=0$ implies that $\tilde{y}_{Ti}$ can be $0$ or $1$. 
Let $\gamma_{Ti}$ indicate that the toxicity outcome $y_{Ti}$ has been ascertained ($\gamma_{Ti}=1$) or is still pending ($\gamma_{Ti}=0$) by the decision time $k$, that is, when the patients in the dose-escalation cohort complete the evaluation period for toxicity, and $u_{Ti}$ ($u_{Ti}\le\tau_{T}$) denote the actual follow-up time for patient $i$ up to that moment. $\tau_{T}$ is defined as the length of the toxicity evaluation window. 
Given the interim toxicity data observed $D^{0}_{Tj}=(\tilde{y}_{T1},...,\tilde{y}_{Tn_{j}},\gamma_{T1},...,\gamma_{Tn_{j}}$) at the current dose level $j$, the joint likelihood function is given by 
\begin{eqnarray*}
L(p_{Tj}|D^{0}_{Tj}) &\propto& \prod_{i=1}^{n_{j}}p_{Tj}^{\gamma_{Ti}y_{Ti}}(1-p_{Tj})^{\gamma_{Ti}(1-y_{Ti})}(1-w_{Ti}p_{Tj})^{1-\gamma_{Ti}}\\
&=& p_{Tj}^{\tilde{x}_{Tj}}(1-p_{Tj})^{m_{Tj}}\prod_{i=1}^{n_{j}}(1-w_{Ti}p_{Tj})^{1-\gamma_{Ti}}\\
&\approx& p_{Tj}^{\tilde{x}_{Tj}}(1-p_{Tj})^{m_{Tj}}\prod_{i=1}^{n_{j}}(1-p_{Tj})^{w_{Ti}(1-\gamma_{Ti})}\\
&=& p_{Tj}^{\tilde{x}_{Tj}}(1-p_{Tj})^{\tilde{m}_{Tj}},\\
\end{eqnarray*}
where $\tilde{x}_{Tj}=\sum_{i=1}^{n_{j}}\gamma_{Ti}y_{Ti}$ is the number of patients who experienced toxicity by the interim time $k$, $\tilde{m}_{Tj}=m_{Tj}+\sum_{i=1}^{n_{j}}w_{Ti}(1-\gamma_{Ti})$ is the effective number of patients who have not experienced toxicity regardless of whether they completed the evaluation,
$m_{Tj}=\sum_{i=1}^{n_{j}}\gamma_{Ti}(1-y_{Ti})$ is the number of patients who have completed the evaluation without experiencing toxicity, and $w_{Ti}=\textup{Pr}(t_{Ti}\le u_{Ti}|y_{Ti}=1)$ can be interpreted as a weight, adjusted for the fact that the toxicity outcome has not been ascertained yet. 
The approximation in the equation for the joint likelihood function is based on the Taylor expansion of $(1-p_{Tj})^{w_{Ti}}$ at $p_{Tj}=0$ by ignoring the second-order and higher terms. 
As suggested by \citet{Lin2020-xx} and \citet{Takeda2020-yc}, we assume that the time-to-toxicity outcome is uniformly distributed over the evaluation period $(0, \tau_{T})$, that is,
$T_{Ti}|y_{Ti}=1\sim \textup{Unif}(0, \tau_{T})$, leading to $w_{Ti}=\textup{Pr}(t_{Ti}\le u_{Ti}|y_{Ti}=1)=u_{Ti}/\tau_{T}$. 
The same approach is applied to the efficacy outcome.

As a result, the likelihoods with pending outcomes are based on the effective toxicity data $\tilde{D}_{Tj}=(\tilde{n}_{Tj},\tilde{x}_{Tj})$ and the effective efficacy data $\tilde{D}_{Ej}=(\tilde{n}_{Ej},\tilde{x}_{Ej})$, where $\tilde{n}_{Tj}$ and $\tilde{n}_{Ej}$ are the effective sample sizes for toxicity and efficacy at dose level $j$, respectively, and $\tilde{x}_{Tj}$ and $\tilde{x}_{Ej}$ are the numbers of patients who experienced toxicity and efficacy by the interim time, respectively. 
Using effective toxicity data $\tilde{D}_{Tj}=(\tilde{n}_{Tj},\tilde{x}_{Tj})$ and effective efficacy data $\tilde{D}_{Ej}=(\tilde{n}_{Ej},\tilde{x}_{Ej})$, $\hat{p}_{Tj}$ and $\hat{p}_{Ej}$ can be replaced by $\tilde{p}_{Tj} =\tilde{x}_{Tj}/\tilde{n}_{Tj}$ and $\tilde{p}_{Ej} =\tilde{x}_{Ej}/\tilde{n}_{Ej}$ as the maximum likelihood estimates of $p_{Tj}$ and $p_{Ej}$, respectively. 
The details of the likelihood with pending outcomes are summarized in Supplementary \ref{sup_likelihood}.

\subsection{Dose-finding algorithm within a subtrial}\label{algorithm}  
Referring to \cite{Takeda2025-do}, the dose allocation decision for the proposed design is guided by the optimal boundaries in Section \ref{boundaries}, the maximum likelihood estimates of $p_{Tj}$ and $p_{Ej}$ (denoted $\tilde{p}_{Tj}$ and $\tilde{p}_{Ej}$) in Section \ref{likelihood}, and the minimum effective dose, $j_{\textup{min}}$, defined by the lowest dose that satisfies $\tilde{p}_{Ej}\ge\lambda_{E}$. $j_{\textup{min}}$ is updated for each dose allocation decision based on the latest $\tilde{p}_{Ej}$.
The dose allocation algorithm is as follows.
\begin{itemize}
\item If $\tilde{p}_{Tj}\le\lambda_{T1}$, 
    \begin{itemize}
    \item the next dose for the dose-escalation cohort is $j+1$,
    \item the admissible dose set for the backfill is $A_{j}=\{j_{\textup{min}},...,j\}$;
    \end{itemize}
\item If $\lambda_{T1}<\tilde{p}_{Tj}\le\lambda_{T2}$, 
    \begin{itemize}
    \item the next dose for the dose-escalation cohort is $j$, 
    \item the admissible dose set for the backfill is $A_{j}=\{j_{\textup{min}},...,j-1\}$;
    \end{itemize}
\item If $\tilde{p}_{Tj}>\lambda_{T2}$, 
    \begin{itemize}
    \item the next dose for the dose-escalation cohort is $j-1$, 
    \item the admissible dose set for the backfill is $A_{j}=\{j_{\textup{min}},...,j-2\}$.
    \end{itemize}
\end{itemize}
The dose-escalation cohort prioritizes the enrollment of patients to explore a higher dose, followed by the backfill cohort that enrolls patients within the admissible dose set $A_j$ until the dose-escalation cohort evaluates sufficient patient outcomes for toxicity to determine the next dose level for the dose-escalation cohort. 
A utility-based enrollment approach is considered for the backfill cohort, that is, the backfill cohort enrolls patients at the dose level that maximizes utility within the admissible dose set $A_j$. To quantify the efficacy-toxicity trade-off, this paper applies the utility proposed by \citet{Zhou2019-nh, Lin2020-vf}:
$$
U_{j}=\sum^{U}_{u=1} \psi_{u}q_{ju},
$$ 
where $\psi_{u}~(u=1,...,U)$ represent utility scores for possible efficacy-toxicity outcomes and $q_{ju}$ denotes their corresponding probabilities at dose level $j$. 
Here, $U=4$ is assumed because the efficacy and toxicity data are summarized as binary outcomes, and $u=1,2,3$, and $4$ correspond to (efficacy, no toxicity), (efficacy, toxicity), (no efficacy, no toxicity), and (no efficacy, toxicity). In this case, the most desirable outcome (no toxicity, efficacy) is assigned the highest score of $\psi_1=100$, and the least desirable outcome (toxicity, no efficacy) is assigned the lowest score of $\psi_4=0$. Given the fixed values of $\psi_1$ and $\psi_4$, the other two outcomes are assigned the scores between $\psi_1$ and $\psi_4$ to reflect clinical desirability.
The efficacy and toxicity probabilities, $p_{Ej}$ and $p_{Tj}$, are assumed to be independent, and the marginal efficacy and toxicity probabilities at dose level $j$ are $q_{j1}+q_{j2}$ and $q_{j2}+q_{j4}$, respectively.
Therefore, we have $q_{j1}=p_{Ej}(1-p_{Tj})$, $q_{j2}=p_{Ej}p_{Tj}$, $q_{j3}=(1-p_{Ej})(1-p_{Tj})$, and $q_{j4}=(1-p_{Ej})p_{Tj}$ at dose level $j$. During the trial, the estimate of mean utility is given by
$$
\tilde{U}_{j}=\sum^{U}_{u=1} \psi_{u}\tilde{q}_{ju}.
$$
This utility framework guides patient enrollment approach in the backfill cohort within subtrial. Further details are provided in Supplementary \ref{sup_utility}.

All patients should complete the toxicity evaluation at the current dose level in the dose escalation cohort.
However, some patients in the backfill cohort may have not completed the evaluation period when the next dose level for the dose-escalation cohort is determined. Patients in the backfill cohort continue to be treated until they complete the evaluation period.
For the dose-escalation cohort, enrollment is stopped when 1) the total number of patients in the dose-escalation cohort reaches the maximum total number of patients in the dose-escalation cohort $N_{esc}$ or 2) the number of patients in the dose-escalation and backfill cohorts at the current dose level reaches the maximum number of patients at a dose level $n_{stop}$ and the same dose level is recommended as the next dose level.
Even if the dose-escalation cohort stops enrollment when the total number of patients in the dose-escalation cohort reaches the maximum $N_{esc}$, the backfill cohort can continue to enroll patients. For the backfill cohort, enrollment is stopped when the total number of patients in the dose-escalation and backfill cohorts at the dose that maximizes the utility reaches the maximum number of patients at a dose level $n_{stop}$.

To avoid allocating patients to ineffective or severely toxic doses, dose elimination criteria are applied before the dose allocation decision when $n_j \ge N_{\textup{ET}}$, where $N_{\textup{ET}}$ is a minimum sample size for dose elimination, for example, 6:
\begin{itemize}
    \item Eliminate dose level $j$ and higher doses if $\textup{Pr}(\tilde{p}_{Tj}>\phi_{T0}|D_{Tj})>C_{T}$, where $C_{T}$ is a probability cutoff for toxicity, for example, 0.95;
    \item Eliminate dose level $j$ if $\textup{Pr}(\tilde{p}_{Ej}<\phi_{E0}|D_{Ej})>C_{E}$, where $C_{E}$ is a probability cutoff for efficacy, for example, 0.95.
\end{itemize}
The trial is terminated if all dose levels are eliminated for any reason. For example, all doses in a subtrial may be eliminated for toxicity before $n_{stop}$ is reached, or efficacy elimination criteria triggered during the subtrial may leave the admissible set empty. The posterior probabilities for efficacy and toxicity can be evaluated based on a beta model for the binary endpoints, assuming that the efficacy and toxicity endpoints follow vague beta priors, for example, Beta(1,1).

This paper adopts the BF-BOIN-ET design~\citep{Takeda2025-do} for dose-finding with backfill witnin a subtrial, though alternative dose-finding designs with backfill~\citep{Dehbi2021-um, Barnett2023-dm, Liu2024-vw, Zhao2024-ng,Chen2025-up} are also applicable within a subtrial.

\subsection{Algorithm for Subtrials}\label{subtrial}  
A trial is considered that aims to identify OBDCs from a matrix of $L$ doses of agent A, denoted as $A_{1}<\ldots<A_{L}$, and $M+1$ doses of agent B, denoted as $B_{0}<\ldots<B_{M}$, where $B_0$ represents no administration of agent B. Agent A is assumed to be a novel anticancer agent under investigation, while agent B is considered an established standard-of-care treatment.
The estimated toxicity and efficacy probabilities at dose combination $(A_{l}, B_{m})$ are denoted by $\hat{p}_{T(l,m)}=y_{T(l,m)}/n_{(l,m)}$ and $\hat{p}_{E(l,m)}=y_{E(l,m)}/n_{(l,m)}$, respectively. Here, $n_{(l,m)}$ represents the number of patients treated, $y_{T(l,m)}$ the number experiencing toxicity, and $y_{E(l,m)}$ the number demonstrating efficacy at this combination.
When pending outcomes are incorporated as described in Section \ref{likelihood}, these estimates $\hat{p}_{T(l,m)}$ and $\hat{p}_{E(l,m)}$ are replaced by their maximum likelihood estimates $\tilde{p}_{T(l,m)}$ and $\tilde{p}_{E(l,m)}$, which are adjusted for patients still under evaluation.

Referring to \citet{Zhang2016-fp} and \citet{Kakizume2024-sr}, the $L \times (M+1)$ dose matrix is divided into subtrials, $S_{k}$, to facilitate both monotherapy and combination therapy dose-finding. Since toxicity remains monotonic within each subtrial by dose level, the dose-finding algorithm introduced in Section \ref{algorithm} is applied in each subtrial to select the next dose combination. 

The following steps and subtrials are considered, and the procedures are summarized in Figure~\ref{fig:dose_adjustment}.
\begin{align*}
    \textup{Step 1:}~&S_{1} = (A_{1}B_{0}, \cdots, A_{L}B_{0}). \\
    \textup{Step 2:}~&S_{2} = (A_{*}B_{1}, \cdots, A_{*}B_{M}, \cdots, A_{\textup{MTD}}B_{M}),\\
&\textup{where}~2.1:~A_{*}=A_{\textup{OBD}}~\textup{if}~\textup{MTD}>\textup{OBD}~\textup{in monotherapy;}\\ 
&~~~~~~~~~~~2.2:~A_{*}=A_{\textup{OBD}-1}~\textup{if}~\textup{MTD}=\textup{OBD}\geq2~\textup{in monotherapy;}\\
&~~~~~~~~~~~2.3:~S_{2} = (A_{\textup{1}}B_{1}, \cdots, A_{1}B_{M})~\textup{if}~\textup{OBD}=\textup{MTD}=1~\textup{in monotherapy.} \\
                    ~&\textup{When}~S_2~\textup{is completed, move to Step 3}~\textup{if}~\hat{p}_{T(*, 1)} \le \lambda_{T1};\\
                    ~&\textup{move to Step 3'}~\textup{if}~\hat{p}_{T(*, 1)} > \lambda_{T1}.\\
    \textup{Step 3:}~&\textup{if}~\hat{p}_{T(*, M-1)} \le \lambda_{T1}, S_{3} = (A_{*+1}B_{M-1}, \cdots, A_{\textup{MTD}}B_{M-1}), \\
                    ~&\textup{if}~\hat{p}_{T(*, M-2)} \le \lambda_{T1}, S_{4} = (A_{*+1}B_{M-2}, \cdots, A_{\textup{MTD}}B_{M-2}), \\
                    & \cdots \\
                    ~&\textup{if}~\hat{p}_{T(*, 1)} \le \lambda_{T1}, S_{M} = (A_{*+1}B_{1}, \cdots, A_{\textup{MTD}}B_{1}). \\
    \textup{Step 3':}~&S_{3}= (A_{*-1}B_{1}, \cdots, A_{*-1}B_{M}), \\
                    &\textup{if}~\hat{p}_{T(*-1, 1)} > \lambda_{T1}, S_{4}= (A_{*-2}B_{1}, \cdots, A_{*-2}B_{M}), \\
                    & \cdots \\
                    &\textup{if}~\hat{p}_{T(2, 1)} > \lambda_{T1}, S_{l}= (A_{1}B_{1}, \cdots, A_{1}B_{M}).
\end{align*}
where $\hat{p}_{T(*, 1)}$ in Step 2 represents the toxicity probability at the dose combination consisting of the starting dose level of agent A and the lowest dose level of agent B, $A_{*+1}$ in Step 3 and $A_{*-1}$ in Step 3' mean one dose level above and below $A_{*}$, respectively.

Step 1 is a monotherapy dose-finding subtrial for agent A. The subtrial $S_{1}$ starts from the lowest dose of monotherapy agent A, $A_{1}B_{0}$. 
Once Step 1 is completed, a candidate OBD of agent A is selected.
For toxicity, an isotonic regression is performed so that the estimated toxicity probability satisfies the monotonicity assumption. Specifically, let $\bar{p}_{T(l,0)}$ denote the isotonic regression estimator of the observed toxicity probability at the dose combination $(l,0)$ using the pooled adjacent violator algorithm (PAVA)~\citep{Barlow-RE-and-Bartholomew-DJ-and-Bremner-JM-and-Brunk-HD1972-wh}. 
The MTD for monotherapy agent A, $A_{\textup{MTD}}B_0$, is selected as the dose whose $\bar{p}_{T(l,0)}$ is closest to the target toxicity probability. 
For efficacy, the observed efficacy probability is used as the estimated efficacy probability such that $\bar{p}_{E(l,m)}=\hat{p}_{E(l,m)}$. The utilities are calculated based on $\bar{p}_{T(l,m)}$ and $\bar{p}_{E(l,m)}$.
Among the dose levels in the subtrial that satisfy tolerability, $A_{l}B_{0} \le A_{\textup{MTD}}B_{0}$, the dose that maximizes the utility in Section~\ref{optimization}, $A_{\textup{OBD}}B_{0}$, is identified as a candidate OBD for agent A.

Steps 2 and 3 are combination therapy dose-finding subtrials for agent A and agent B. 
Once a dose $A_{\textup{OBD}}B_{0}$ is selected as the candidate OBD for agent A, Step 2 is initiated to identify the candidate OBDC of agent A and agent B in a subtrial. The subtrial $S_{2}$ starts from the dose combination consisting the candidate OBD of monotherapy agent A and the lowest dose level of agent B, $A_{\textup{OBD}}B_{1}$.   
However, when $\textup{OBD}=\textup{MTD}$ and $\textup{OBD}\geq2$, the initial dose level for agent $A$ in $S_2$ is defined as $A_{\textup{OBD}-1}B_{1}$ due to safety concerns in combination therapy. 
Once a dose combination $A_{l^*}B_{m^*}$ is selected as the candidate OBDC in the subtrial $S_{2}$, the subsequent subtrial $S_{3}$ starts to search the higher dose combination space if the lowest dose combination is tolerable, that is, $\hat{p}_{T(*, 1)}\le\lambda_{T1}$. 
Otherwise, Step 3' is initiated to search lower dose combination spaces until a candidate OBDC is detected.

Within each subtrial, the toxicity probability is assumed to increase monotonically with the dose levels. The subtrials are conducted according to the following steps:
\begin{enumerate}
\item[(i)] Enroll $n$ patients and observe toxicity outcomes. Then, the next dose combination is determined based on the dose assignment algorithm introduced in Section \ref{algorithm}. Stop the subtrial when the average total number of patients treated at the dose combination selected as the next dose reaches a predefined fixed number of patients, $n_{stop}$, and move to the next subtrial. The selected dose combination is called the candidate OBDC in the subtrial.

\item[(ii)] Repeat step (i) until all subtrials are completed. 
\end{enumerate}

\subsection{Dose optimization}\label{optimization} 
Once the study is completed, the final OBDCs are selected among the admissible dose combinations, including agent A monotherapy $A_{l}B_{0}$. It is assumed that the toxicity probability increases monotonically with the dose level. In order to select the admissible dose combinations, MTD combinations (MTDCs) are selected for each agent B dose level. The MTDC is defined as the dose combination with the toxicity probability closest to the target toxicity probability $\phi_{T0}$. The bivariate isotonic regression model is applied to estimate the toxicity probability for each dose combination while satisfying the monotonicity assumption. Before applying the bivariate isotonic regression model, the toxicity probability of each dose combination, $\pi_{T(l,m)}$, is calculated using the following Beta-Binomial model because some dose combinations may not have been administered during the study,
$$
  x_{T(l,m)}|\pi_{T(l,m)} \sim \textup{Binomial}(\pi_{T(l,m)}, n_{(l,m)}), \; \pi_{T(l,m)} \sim \textup{Beta}(0.05, 0.05).
$$
As a result, $\pi_{T(l,m)} = (x_{T(l,m)}+0.05)/(n_{(l,m)}+0.1)$. Then, the toxicity probability of ($A_{l}$, $B_{m}$), $\bar{p}_{T(l,m)}$, is estimated based on the two-dimensional PAVA.
The MTDC of each agent B dose level $m$, $d_{\textup{MTDC}, m}$, is then selected as the dose combination that has the estimated toxicity probability closest to the target toxicity rate $\phi$, that is,
$$
    d_{\textup{MTDC}, m} = \underset{l \in (1, \cdots, L)} {\operatorname{\arg \min}} |\bar{p}_{T(l,m)}-\phi_{T0}|, \quad (m = 0, 1, \cdots, M)
$$
The admissible dose combinations are all dose combinations that do not exceed $d_{\textup{MTDC}, m}$. For example, in case of $(3\times2)$ dose combinations, if $d_{\textup{MTDC}} = \{d_{\textup{MTDC}, 0}, d_{\textup{MTDC}, 1}\} = \{(A_3, B_0), (A_2, B_1)\}$, the admissible dose combinations are $\{(A_1, B_0), (A_2, B_0), (A_3, B_0),$\\$ (A_1, B_1), (A_2, B_1)\}$. 
For efficacy, the observed efficacy probability is used as the estimated efficacy probability such that $\bar{p}_{E(l,m)}=\hat{p}_{E(l,m)}$.

Based on the estimated toxicity and efficacy probabilities, $\bar{p}_{(l,m)}$ and $\bar{p}_{E(l,m)}$, and the efficacy-toxicity trade-off utility introduced in Section \ref{algorithm} and Supplementary~\ref{sup_utility}, the estimate of mean utility when the trial is completed is given by
$$
d_{\textup{OBDC}} = \mathop{\arg \max}_{(A_{l}, B_{m}) \in AD} \bar{U}(A_{l}, B_{m})=\mathop{\arg \max}_{(A_{l}, B_{m}) \in AD} \sum^{U}_{u=1} \psi_{u}\bar{q}_{(l,m)u},
$$ 
where AD is the set of admissible dose combinations, $\psi_{u}~(u=1,...,U)$ represents utility scores corresponding to the possible efficacy-toxicity outcomes, $\bar{q}_{(l,m)u}$ denotes the respective probabilities of observing the outcomes at dose combination $(A_{l}, B_{m})$, and $\bar{q}_{(l,m)1}=\bar{p}_{E(l,m)}(1-\bar{p}_{T(l,m)})$, $\bar{q}_{(l,m)2}=\bar{p}_{E(l,m)}\bar{p}_{T(l,m)}$, $\bar{q}_{(l,m)3}=(1-\bar{p}_{E(l,m)})(1-\bar{p}_{T(l,m)})$, and $\bar{q}_{(l,m)4}=(1-\bar{p}_{E(l,m)})\bar{p}_{T(l,m)}$ at dose combination $(A_{l}, B_{m})$ when $U=4$.

\section{Simulation studies}\label{simulation}
The operating characteristics of the proposed design were evaluated via simulation studies, with comparisons made to existing designs that integrated separate monotherapy and combination therapy approaches.

\subsection{Simulation study design}\label{simulationdesign} 
The design described below applied our proposed method.
\begin{itemize}
    \item[(i)] the proposed design, denoted BOIN-MC-1
    \item[(ii)] the proposed design without backfill, denoted BOIN-MC-2
\end{itemize}
To our knowledge, no existing designs genuinely targeted the same design problem for seamless monotherapy and combination therapy optimization. Therefore, as benchmarking against constituent designs, we included: a design combining the BOIN design~\citep{Liu2015-pc} for monotherapy with the waterfall design~\citep{Zhang2016-fp} for combination therapy; a design integrating the BOIN-ET design~\cite{Takeda2018-rs} for monotherapy with the BOIN-ETC design~\citep{Kakizume2024-sr} for combination therapy; and a design combining the BOIN12 design~\citep{Lin2020-vf} for monotherapy with the Comb-BOIN12 design~\citep{Lu2025-fb} for combination therapy.
Two versions were generated for each design by modifying the combination therapy starting dose.
\begin{itemize}
    \item[(iii)] the BOIN design for monotherapy, followed by the waterfall design for combination therapy with starting dose $(A_1, B_1)$, denoted BOIN-W-1
    \item[(iv)] the BOIN design for monotherapy, followed by the waterfall design for combination therapy with starting dose $(A_{\textup{MTD}-1}, B_1)$, where $A_{\textup{MTD}-1}$ is one dose level below the MTD for monotherapy, denoted BOIN-W-2
    \item[(v)] the BOIN-ET design for monotherapy, followed by the BOIN-ETC design for combination therapy with starting dose $(A_1, B_1)$, denoted BOIN-ET-1
    \item[(vi)] the BOIN-ET design for monotherapy, followed by the BOIN-ETC design for combination therapy with starting dose $(A_{\textup{OBD}}, B_1)$, where $A_{\textup{OBD}}$ is the OBD for monotherapy, denoted BOIN-ET-2
    \item[(vii)] the BOIN12 design for monotherapy, followed by the Comb-BOIN12 design for combination therapy with starting dose $(A_1, B_1)$, denoted BOIN12-Combo-1
    \item[(viii)] the BOIN12 design for monotherapy, followed by the Comb-BOIN12 design for combination therapy with starting dose $(A_{\textup{OBD}}, B_1)$, where $A_{\textup{OBD}}$ is the OBD for monotherapy, denoted BOIN12-Combo-2
\end{itemize}
In designs (ii) through (viii), no backfill cohorts were incorporated. In designs (iii) through (viii), the MTD and OBD were identified independently for the monotherapy and combination therapy, respectively.
The target toxicity probability $\phi_{T0}=0.30$ was adopted for all designs. 
For the BOIN-MC-1 and BOIN-MC-2 designs, the minimum required efficacy probability $\phi_{E0}=0.25$ was adopted, and the design parameters were $\phi_{T1}=0.6\phi_{0}$, $\phi_{T2}=1.4\phi_{0}$, and $\phi_{E1}=0.6\phi_{E0}$ as in Supplementary~\ref{sup_boundaries}. As a result, the optimal values of ($\lambda_{T1}$, $\lambda_{T2}$, $\lambda_{E}$) for the probability of toxicity and efficacy were obtained as $\lambda_{T1}=0.236$, $\lambda_{T2}=0.359$, and $\lambda_{E}=0.197$.
For the BOIN-W-1 and BOIN-W-2 designs, the design parameters were $\phi_{T1}=0.6\phi_{T0}$ and $\phi_{T2}=1.4\phi_{T0}$ the same as the BOIN-MC-1 and BOIN-MC-2 designs. As a result, the optimal values of ($\lambda_{T1}$, $\lambda_{T2}$) for the probability of toxicity were obtained as $\lambda_{1}=0.236$ and $\lambda_{2}=0.359$.
For the BOIN-ET-1 and BOIN-ET-2 designs, the target efficacy probability $\phi_{E0}=0.50$ was adopted. The design parameters were $\phi_{T1}=0.1\phi_{T0}$, $\phi_{T2}=1.4\phi_{T0}$, and $\phi_{E1}=0.6\phi_{E0}$ as in the original paper~\citep{Takeda2018-rs}. As a result, the optimal values of ($\lambda_{T1}$, $\lambda_{T2}$, $\lambda_{E}$) for the toxicity probability and the efficacy probability were obtained to be $\lambda_{1}=0.130$, $\lambda_{2}=0.351$, and $\lambda_{E}=0.392$.

The maximum number of patients in a subtrial was $N_{esc}=30$, the maximum number of patients at each dose level was $n_{stop}=12$, and the cohort size was 3. Investigators often set the maximum number of patients as the number of dose levels $\times$ 6 patients (2 cohorts) in real clinical trials. Following this approach, the maximum number of patients in a subtrial was 5 dose levels (the maximum number of dose levels in a subtrial) $\times$ 6 patients $=$ 30 patients. 
In the BOIN-MC-1 design, no cohort size was defined in the backfill cohort. Skipping doses was not allowed in each subtrial from a safety point of view. The early termination criteria for efficacy and toxicity described in Section \ref{algorithm} were applied to all designs. 
Dose-finding trials were simulated 10,000 times. With 10,000 simulations, the maximum Monte Carlo standard error for estimated binomial operating characteristics, such as selection percentages and early stopping percentages, was 0.005. The empirical standard deviation of the total trial duration (months) was presented in Table \ref{tab:sd_total_duration_all_part}. In addition, the empirical standard deviation of the total number of patients was presented in Table \ref{tab:sd_total_n_all_part}.

For all designs, when the trial was completed, the dose combinations that maximized utility ware selected from the admissible dose combinations that satisfied tolerability as the OBDCs, as described in Section \ref{optimization}. Following \citet{Lin2020-vf}, the utilities were set as (efficacy, no toxicity)=100, (no efficacy, toxicity)=0, (efficacy, toxicity)=60, and (no efficacy, no toxicity)=40.

The evaluation periods (in days) for toxicity and efficacy were $\tau_{T}=28$ and $\tau_{E}=56$, as the efficacy evaluation period is typically longer than the toxicity evaluation period in oncology dose-finding studies. The accrual rate was one patient per 14 days. The time-to-toxicity outcomes were simulated from a Weibull distribution to ensure that
50\% of the toxicity and efficacy outcomes occurred in the second half of the evaluation period. To generate toxicity and efficacy outcomes, a correlation between toxicity and efficacy probability was also considered to be $0.2$ using the Gaussian copula. The details of data generation were summarized in Supplementary~\ref{sup_weibull}.

The true toxicity and efficacy probabilities for each dose combination were summarized in Table \ref{tab:tox_eff_scenarios}. Agent A was assumed to be a novel anticancer agent under investigation, while agent B was considered an established standard-of-care treatment that had limited, well-established dose options. In all scenarios, toxicity probabilities increased with dose levels. The dose-efficacy relationships exhibited either increasing or plateau patterns. 
The bold numbers indicate the dose level that was identified as the OBDC across both monotherapy and combination therapy. All OBDCs were tolerable with sufficient efficacy. The red asterisk marks the dose level considered to be the OBD for monotherapy in scenarios where no suitable OBDC could be identified in the combination therapy setting.
The R code for the BOIN-MC design was available at GitHub: \url{https://github.com/masahikoji/BOIN-MC.git}.

\subsection{Simulation results}\label{simulationresults}
\begin{figure}[tbp]\centering
\includegraphics[width=\linewidth]{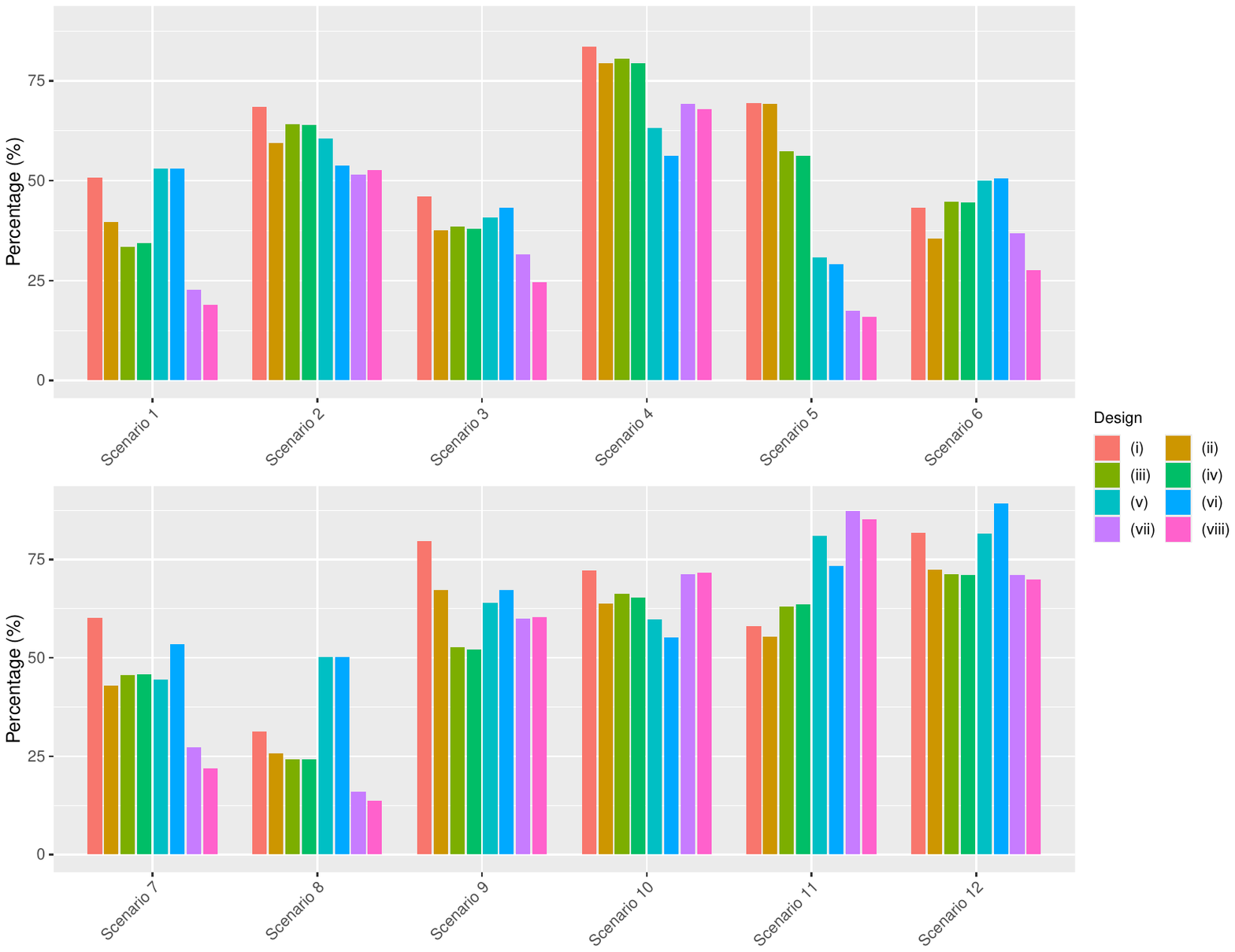}
\caption{Percentage of correct OBDC selection across all part. (i) BOIN-MC-1, (ii) BOIN-MC-2, (iii) BOIN-W-1, (iv) BOIN-W-2, (v) BOIN-ET-1, (vi) BOIN-ET-2, (vii) BOIN12-Combo-1, and (viii) BOIN12-Combo-2.}
\label{fig:correct_obd_percent}
\end{figure}

\begin{figure}[tbp]\centering
\includegraphics[width=\linewidth]{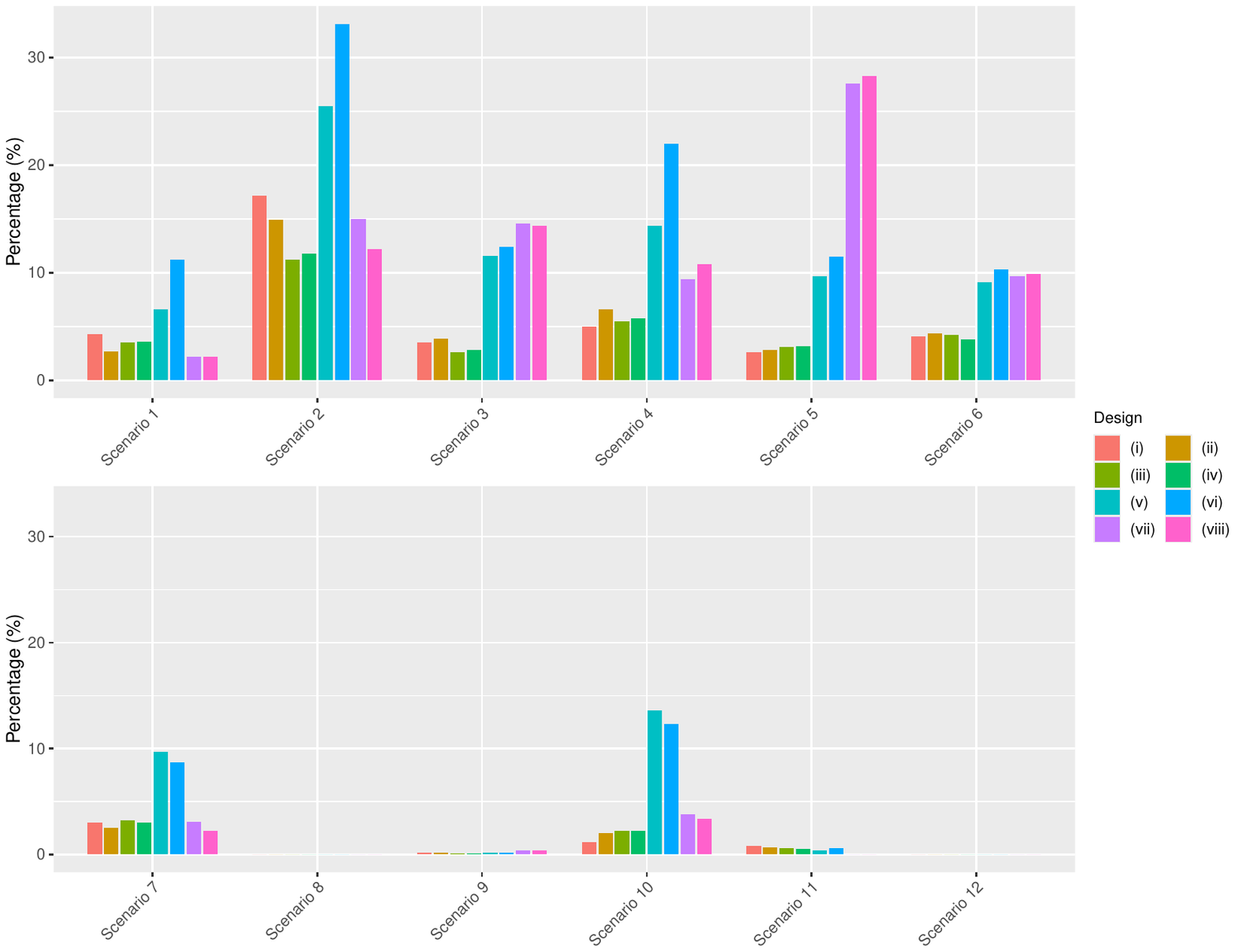}
\caption{Percentage of overdosing selection as OBDC across all part. (i) BOIN-MC-1, (ii) BOIN-MC-2, (iii) BOIN-W-1, (iv) BOIN-W-2, (v) BOIN-ET-1, (vi) BOIN-ET-2, (vii) BOIN12-Combo-1, and (viii) BOIN12-Combo-2.}
\label{fig:od_obd_percent}
\end{figure}

\begin{figure}[tbp]\centering
\includegraphics[width=\linewidth]{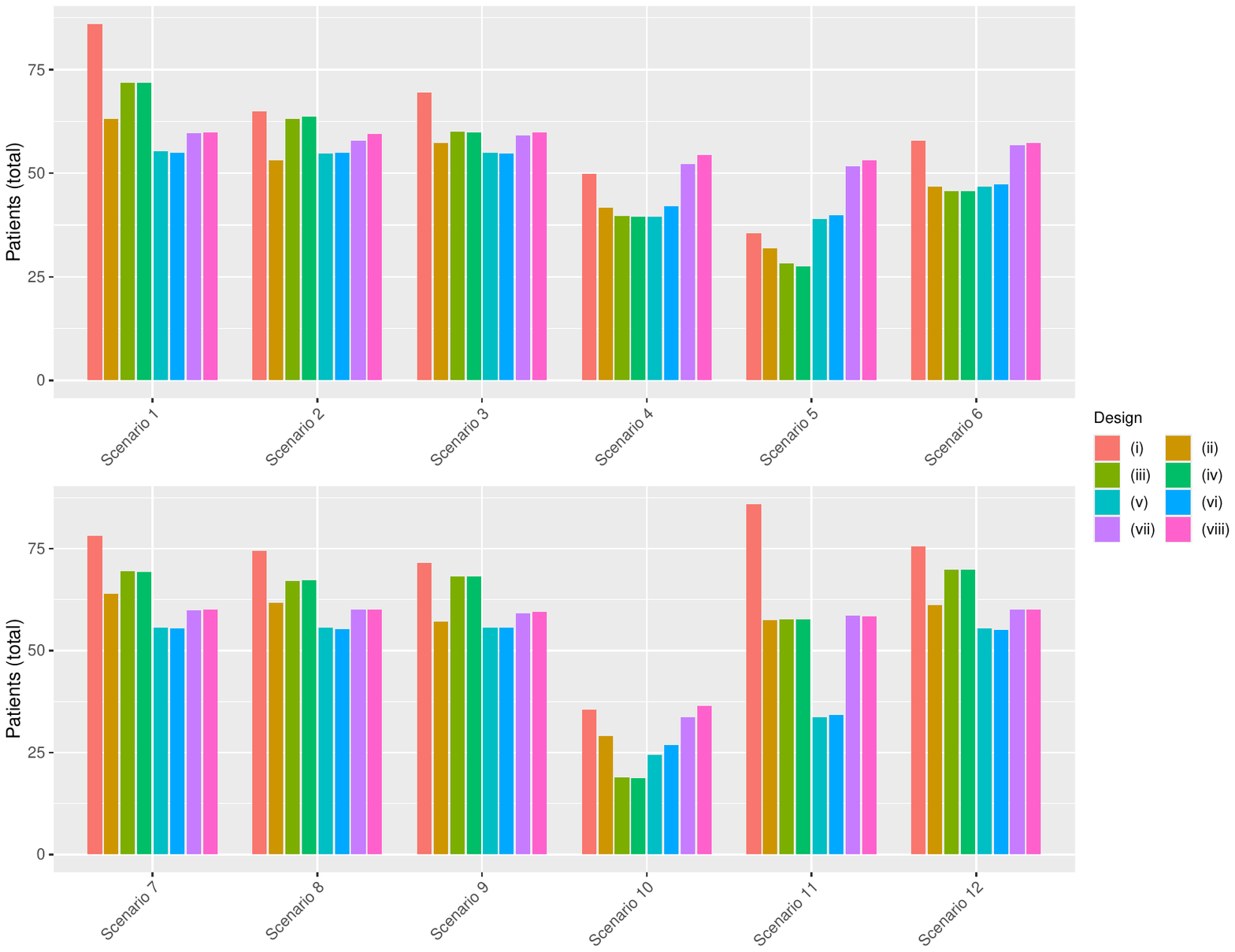}
\caption{Average total number of patients across all part. (i) BOIN-MC-1, (ii) BOIN-MC-2, (iii) BOIN-W-1, (iv) BOIN-W-2, (v) BOIN-ET-1, (vi) BOIN-ET-2, (vii) BOIN12-Combo-1, and (viii) BOIN12-Combo-2.}
\label{fig:total_patients}
\end{figure}

\begin{figure}[tbp]\centering
\includegraphics[width=\linewidth]{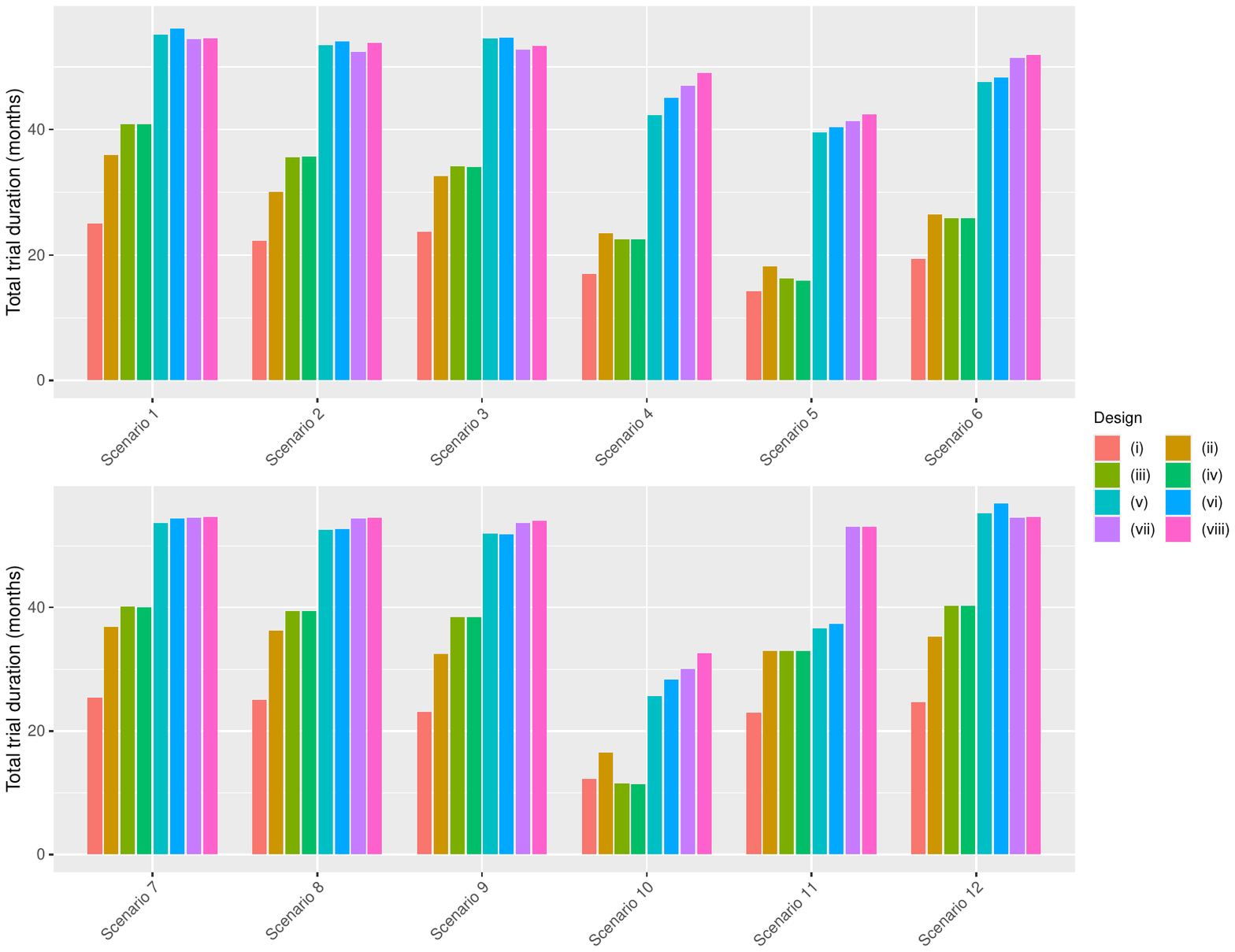}
\caption{Average total trial duration across all part. (i) BOIN-MC-1, (ii) BOIN-MC-2, (iii) BOIN-W-1, (iv) BOIN-W-2, (v) BOIN-ET-1, (vi) BOIN-ET-2, (vii) BOIN12-Combo-1, and (viii) BOIN12-Combo-2.}
\label{fig:trial_duration}
\end{figure}

The operating characteristics of the designs, organized by scenarios, were summarized across all trial components as follows: percentage of correct OBD selection in Figure \ref{fig:correct_obd_percent} and Table \ref{tab:correct_obd_percent}, percentage of overdosing selections in Figure \ref{fig:od_obd_percent} and Table \ref{tab:od_obd_percent}, average total patient enrollment in Figure \ref{fig:total_patients} and Table \ref{tab:total_alln_rawsum}, and average trial duration in Figure \ref{fig:trial_duration} and Table \ref{tab:total.duration}.
Additionally, supplementary tables presented the following: percentage of OBDC selection by dose level in Tables \ref{tab:obd_percent_scenario1}-\ref{tab:obd_percent_scenario8}, average number of patients by dose level in Tables \ref{tab:ave_alln_scenario1}-\ref{tab:ave_alln_scenario12}, percentage of MTD selection in monotherapy in Table \ref{tab:mono_correct_mtd_percent}, 
percentage of MTDC selection in combination therapy in Table \ref{tab:comb_correct_mtd_percent}, and percentage of OBDC selection in combination therapy in Table \ref{tab:comb_correct_obd_percent}.
The probabilities of edge-case stopping events were summarized in Table~\ref{tab:edge-case-stop-probability}.
The results of average number of patients treated above the correct OBDC across all part were shown in Table \ref{tab:n_above_true_mtdc}. The results of percentage of trials terminated early for toxicity across all part were shown in Table \ref{tab:earlystop_tox_percent}. 
The operating characteristics for the small sample sizes (n=3, 6) were in Tables \ref{tab:toxicity-decision-probability-n3} and \ref{tab:toxicity-decision-probability-n6}. 

Under Scenarios 1, 2, and 4, the OBDCs were in both monotherapy and combination therapy. In these scenarios, the BOIN-MC-1 design demonstrated the highest percentage of correct OBDC selection, ranging from 50.8\% to 83.6\%. The performance of other designs varied in each scenario. When the OBDCs were located at lower dose levels, the BOIN-MC-2, BOIN-W-1, BOIN-W-2 designs showed the second-best performance. When the OBDCs were located at higher dose levels, the BOIN-ET-1 and BOIN-ET-2 designs showed the second-best performance. 
Under Scenarios 3 and 7, the OBDCs were in only combination therapy. In these scenarios, the BOIN-MC-1 design showed the highest percentage of correct OBDC selection, ranging from 46.0\% to 60.1\%. The performance of other designs varied in each scenario. 
Under Scenarios 5, 6, and 8, only one OBDC was in combination therapy. In Scenario 5, the BOIN-MC-1 and BOIN-MC-2 designs provided the highest percentage of correct OBDC selection, followed by the BOIN-W-1 and BOIN-W-2 designs. In Scenario 6, the BOIN-MC-1, BOIN-W-1, BOIN-W-2, BOIN-ET-1, and BOIN-ET-2 designs showed similar performance. In Scenario 8, the BOIN-ET-1 and BOIN-ET-2 designs showed the highest percentage of correct OBDC selection, followed by the BOIN-MC-1 design. 
Under Scenarios 9, the OBDCs were in both monotherapy and combination therapy with non-monotone efficacy pattern. In this scenario, the BOIN-MC-1 design demonstrated the highest percentage of correct OBDC selection, followed by the BOIN-MC-2 design. 
Under Scenario 10, only one OBDC was in combination therapy because all combination therapies were toxic (super-additive combination toxicity scenario). In this scenario, the BOIN-MC-1, BOIN12-Combo-1, and BOIN12-Combo-2 designs provided a higher percentage of correct OBDC selection. 
Under Scenario 11, the OBDCs were in both monotherapy and combination therapy with flat-efficacy plateau profiles. In this scenario, the BOIN-ET-1, BOIN12-Combo-1, and BOIN12-Combo-2 designs showed correct OBDC selection percentages higher than 80\%. 
Under Scenario 12, the OBDCs were at matrix boundaries in both monotherapy and combination therapy. In this scenario, the BOIN-MC-1 and BOIN-ET-2 designs provided correct OBDC selection percentages higher than 80\%.

The BOIN-MC-1 design demonstrated a similar percentage of overdosing selection as OBDC as the BOIN-MC-2, BOIN-W-1, and BOIN-W-2 designs across all scenarios except Scenario 2. The BOIN-ET-1 and BOIN-ET-2 designs exhibited a slightly higher percentage of overdosing selection as OBDC across scenarios. The BOIN12-Combo-1 and BOIN12-Combo-2 designs showed a higher percentage of overdosing selection as OBDC in Scenarios 2, 3, 4, 5, and 6.
In all scenarios, the BOIN-MC-1, BOIN-MC-2, BOIN-W-1, and BOIN-W-2 designs similarly allocated patients to doses above the true MTDC across all parts. In many scenarios, the BOIN-MC-1 and BOIN-MC-2 designs differently allocated patients to doses above the true MTDC across all parts from the BOIN-ET-1, BOIN-ET-2, BOIN12-Combo-1, and BOIN12-Combo-2 designs.

The BOIN-MC-1 design enrolled more patients than other designs due to backfill patients, for example, an average of 1.7 to 14.6 more patients than the BOIN-W-1 design. However, the backfill approach enabled concurrent patient enrollment; as a result, the BOIN-MC-1 design significantly shortened the average total trial duration across all scenarios. The BOIN-MC-2 design saved the average total number of patients. The BOIN-ET-1, BOIN-ET-2, BOIN12-Combo-1, and BOIN12-Combo-2 designs also reduced the average total number of patients. On the other hand, the BOIN-ET-1, BOIN-ET-2 BOIN12-Combo-1, and BOIN12-Combo-2 designs required a longer total trial duration because they had to wait for the efficacy evaluation to determine the next dose level.

In all scenarios, the BOIN-MC-1 and BOIN-MC-2 designs similarly terminated early. In scenarios 1, 2, 7, and 9, the BOIN-MC-1 and BOIN-MC-2 designs terminated earlier than the other designs.
The BOIN-MC-1 and BOIN-MC-2 designs terminated the enrollment when 
(a) the dose-escalation cohort reached $N_{esc}$ but the backfill cohort had substantially fewer patients than $n_{stop}$; (b) all doses in a subtrial were eliminated for toxicity before $n_{stop}$ was reached; or (c) efficacy elimination criteria were triggered mid-subtrial in ways that left the admissible set empty. The BOIN-MC-1 rarely terminated due the reason (a) and the BOIN-MC-1 and BOIN-MC-2 designs showed similar stopping patterns due to reasons (b) and (c). 

In summary, the proposed BOIN-MC-1 design provided the highest or comparable percentage of correct OBDC selection across all scenarios except Scenarios 8 and 11. The performance of other designs varied in each scenario. Generally, the BOIN-MC-2, BOIN-W-1, and BOIN-W-2 designs worked well when the OBDCs were located at lower dose levels. In contrast, the BOIN-ET-1 and BOIN-ET-2 designs provided good performance when the OBDCs were located at higher dose levels. The performance of the BOIN12-Combo-1 and BOIN12-Combo-2 designs varied across scenarios.
The BOIN-MC-1 design provided good overdosing control and, despite enrolling more patients, significantly reduced the average total trial duration across all scenarios.


\subsection{Sensitivity analysis}\label{sensitivity}
The results of the sensitivity analysis using different utility scores were shown in Table \ref{tab:correct-obd-percent-compare}. The different utility settings provided similar correct OBD selection percentages, demonstrating the robustness of the utility function.

The results of the sensitivity analysis using different correlation values were shown in Table \ref{tab:correct-obd-percent-compare-correlation}.  The different correlation values provided similar correct OBD selection percentages, demonstrating limited influence of correlation on the results.

The results of the sensitivity analysis with the combination starting dose set at (1,1) were shown in Table \ref{tab:correct-obd-percent-compare-dose}.  The two starting dose rules for combination subtrials provided similar correct OBD selection percentages.

The results of the sensitivity analysis with dose matrices of 4 × 3 and 5 × 4 were shown in Tables \ref{tab:tox_eff_scenarios_diff} and \ref{tab:correct_obd_percent_diff_mat}. Under these scenarios, the BOIN-MC-1 design provided higher or comparable correct OBD selection percentages compared with other designs, demonstrating robustness to varying dose matrix sizes.

\section{Discussion}\label{discussion}
We have proposed a seamless dose-optimization design for monotherapy and combination therapy based on efficacy and toxicity outcomes, considering real-world clinical practice, where investigators commonly conduct dose-finding trials for novel anticancer agents before exploring combinations with standard agents that have limited,
well-established dose options. The proposed design considers adaptive subtrials for monotherapy and combination therapy to efficiently search for OBDCs, and enables patient backfilling at lower dose levels while maintaining focus on dose escalation to explore higher dose levels. Our simulations show that the proposed design could outperform or be comparable to other available approaches in terms of the percentage of correct OBDC selection. In addition, compared with the BOIN-ET-1 and BOIN-ET-2 designs, which, like our proposed approach, incorporate both safety and efficacy, the proposed design demonstrated a lower probability of selecting overdosing levels. Although the application of backfilling cohorts resulted in an 
increase in the average total number of treatment allocations relative to the other designs, the capacity to address pending outcomes enabled the overall trial duration to be reduced.

This paper focuses on seamless dose-optimization for both monotherapy and combination therapy. However, the proposed approach can also be applied exclusively to combination therapy dose-optimization following the completion of monotherapy dose-escalation.
Additionally, while this paper assumes that subtrials open sequentially after the previous subtrial completes, the framework can readily accommodate staggered subtrial initiation rather than strictly sequential manner. This flexibility proves particularly valuable for accelerating dose-optimization across monotherapy and combination therapies in clinical practice. Backfilling plays a crucial role in this context, as it generates early safety and efficacy data that enable timely initiation of subsequent subtrials.

Similar to other dose-finding trial designs, the proposed design applies dose elimination criteria based on posterior probabilities to avoid allocating patients to ineffective or severely toxic doses. On this basis, as \citet{Pin2024-hi} suggests, an adaptive weighting approach could be beneficial for adjusting the probability of assignment as an alternative to elimination criteria based on posterior probabilities.

Although we apply the utility function to measure the efficacy-toxicity trade-off for identifying the OBD, alternative approaches can also be applied to identify the OBD. \citet{Yamaguchi2024-lu} reported the operating characteristics of various OBD selection approaches through comprehensive simulation studies.

This paper assumes independence of the efficacy and toxicity outcomes to estimate the optimal boundaries because the correlation between efficacy and toxicity outcomes is not always predicted at the design stage, especially in novel anticancer agents. Previous research has also shown that ignoring the correlation between efficacy and toxicity has little impact on the performance of the design \citep{Cai2014-kj}. Under this situation, the independence assumption between the efficacy and toxicity outcomes could be reasonable. However, the assumption may not always be appropriate. Therefore, it may be valuable to consider jointly modeling efficacy and toxicity outcomes.

The BOIN-MC starts Step 2 at $A_{\textup{OBD}-1} B_{1}$ when MTD = OBD $\ge$ 2 in monotherapy for safety because regulatory agencies usually recommend that combination therapy dose-finding should start at a lower dose than the monotherapy MTD. However, this procedure could be overly conservative or insufficiently conservative depending on the degree of pharmacokinetic or pharmacodynamic interaction between the agents. It would be valuable to consider alternative algorithms to determine the starting dose for combination therapy based on monotherapy information and pharmacokinetic or pharmacodynamic interaction between the agents.

The BOIN-MC design uses completed subtrials' information to define subsequent subtrials and establish starting doses, and no information sharing occurs across subtrials for dose-finding decisions even though the BOIN-MC design uses all the information from all subtrials to identify the OBDC when the study is completed. This no borrowing approach is sufficient to identify the OBDC because other model-assisted designs~\citep{Zhang2016-fp, Lin2017-cf, Kakizume2024-sr, Lu2025-fb} for combination therapy adopt the same strategy and our simulation results demonstrate sufficient performance. Nevertheless, it could be valuable extension of the BOIN-MC to consider borrowing information across subtrials during the dose-finding trial. 

The current admissibility criteria in the BOIN-MC design, which restrict dose combinations to those not exceeding the MTDC, may be overly stringent by focusing exclusively on toxicity. Incorporating both toxicity and efficacy thresholds via Bayesian posterior probabilities could offer a more balanced approach to defining admissible dose combinations.

The BOIN-MC design simultaneously handles: (1) toxicity and efficacy outcomes, (2) different evaluation windows, (3) completed and pending patients, (4) dose-escalation and backfill cohorts, and (5) multiple dose levels across multiple subtrials. Implementing BOIN-MC in real clinical trials requires the drug development team to address operational issues. Therefore, physicians, biostatisticians, and other stakeholders must collaborate closely to implement this novel design for efficient drug development. On this basis, the average number of concurrently active dose levels is ranged from 1.21 to 1.56 across scenarios as in Table~\ref{tab:ave_concurrent_dose}. Therefore, the BOIN-MC design should be manageable through development team collaboration.

The BOIN-MC design sequentially evaluates multiple dose combinations, uses multiple elimination criteria, and selects a single OBDC from a set of admissible combinations. As a result, BOIN-MC may have a multiplicity issue. Therefore, investigators should carefully confirm the operating characteristics, including the false-positive OBDC selection rate, through simulation at the design stage.

When implementing the BOIN-MC design, investigators should consider alignment with regulatory authorities and other relevant guidelines. Ensuring alignment with regulatory authorities remains a limitation of the BOIN-MC design implementation. Details are discussed in Supplementary \ref{sup_authority}.

The proposed design can be extended in several ways. First, the BOIN-MC design can integrate clinical pharmacokinetic, pharmacodynamic, and pharmacogenomic outcomes into the algorithm and decision-making process such as~\citep{Takeda2023-ka,Xia2024-lb,Sun2025-my}. Second, it would be a valuable extension to incorporate a randomization stratum for dose comparison, similar to approaches proposed by~\cite{Zhao2025-hp,Chen2025-up,Takeda2026-uq}. 
Although the BOIN-MC design focuses on handling binary endpoints of efficacy and toxicity, it can be easily extended to handle quasi-binary endpoints, such as in the generalized BOIN designs~\citep{Mu2019-fk,Takeda2022-uz,Takeda2022-ec,Takeda2023-ad}.
Furthermore, given that the number of patients in oncology dose-finding trials is often limited, it would be valuable to consider incorporating historical information into the proposed design, similar to the iBOIN designs~\citep{Zhou2021-hu, Zhao2023-gf}. It could also be valuable to incorporate monotherapy subtrial information into the boundary formulation for combination trials.
Finally, while this paper adopts an algorithm to adaptively define subtrials, it would be worth considering alternative algorithms for defining subtrials, such as those used in the BOIN-ETC design~\citep{Kakizume2024-sr}.


\subsection*{Acknowledgments}
The authors wish to thank the editors and the reviewers for their thoughtful and constructive comments and suggestions. For MK, this work was supported by JSPS KAKENHI (Grant Number JP26K21185).

\subsection*{Conflict of interest}
The authors have declared no conflict of interest.

\subsection*{Data Availability Statement}
The data that support the findings of this study are available from the corresponding author upon reasonable request.

\bibliographystyle{abbrvnat}
\bibliography{paperpile}%

\begin{table}[htb]
\centering
\caption{True toxicity and efficacy probabilities. The probabilities at OBDCs across both monotherapy and combination therapy were in boldface with dagger.}
\label{tab:tox_eff_scenarios}
\resizebox{0.93\textwidth}{!}{%
\begin{tabular}{cc}
\toprule
\multicolumn{1}{c}{\bf Scenario 1} & \multicolumn{1}{c}{\bf Scenario 2}\\
\midrule
\begin{tabular}{lccc}
     & \textbf{B0} & \textbf{B1} & \textbf{B2} \\
\midrule
A1 & (0.01,\,0.10) & (0.05,\,0.20) & (0.07,\,0.30) \\
A2 & (0.04,\,0.20) & (0.10,\,0.30) & (0.13,\,0.40) \\
A3 & (0.11,\,0.30) & (0.15,\,0.40) & \textbf{(0.30,\,0.60)}$^{\dagger}$ \\
A4 & (0.15,\,0.40) & \textbf{(0.30,\,0.60)}$^{\dagger}$ & (0.45,\,0.65) \\
A5 & \textbf{(0.30,\,0.60)}$^{\dagger}$ & (0.45,\,0.65) & (0.55,\,0.70) \\
\end{tabular}
 & 
\begin{tabular}{lccc}
     & \textbf{B0} & \textbf{B1} & \textbf{B2} \\
\midrule
A1 & (0.05,\,0.05) & (0.15,\,0.15) & \textbf{(0.25,\,0.45)}$^{\dagger}$ \\
A2 & (0.10,\,0.10) & \textbf{(0.25,\,0.45)}$^{\dagger}$ & (0.45,\,0.50) \\
A3 & (0.15,\,0.20) & (0.45,\,0.50) & (0.55,\,0.55) \\
A4 & \textbf{(0.25,\,0.45)}$^{\dagger}$ & (0.55,\,0.55) & (0.65,\,0.60) \\
A5 & (0.45,\,0.50) & (0.65,\,0.60) & (0.75,\,0.65) \\
\end{tabular}
 \\[1em]
\midrule
\multicolumn{1}{c}{\bf Scenario 3} & \multicolumn{1}{c}{\bf Scenario 4}\\
\midrule
\begin{tabular}{lccc}
     & \textbf{B0} & \textbf{B1} & \textbf{B2} \\
\midrule
A1 & (0.10,\,0.10) & (0.10,\,0.30) & (0.15,\,0.40) \\
A2 & (0.20,\,0.20) & (0.15,\,0.40) & \textbf{(0.25,\,0.60)}$^{\dagger}$ \\
A3 & (0.30,\,0.30) & \textbf{(0.25,\,0.60)}$^{\dagger}$ & (0.45,\,0.60) \\
A4 & (0.45,\,0.40) & (0.50,\,0.60) & (0.60,\,0.60) \\
A5 & (0.55,\,0.60) & (0.60,\,0.60) & (0.70,\,0.60) \\
\end{tabular}
 & 
\begin{tabular}{lccc}
     & \textbf{B0} & \textbf{B1} & \textbf{B2} \\
\midrule
A1 & (0.10,\,0.20) & \textbf{(0.30,\,0.60)}$^{\dagger}$ & (0.50,\,0.60) \\
A2 & \textbf{(0.30,\,0.60)}$^{\dagger}$ & (0.45,\,0.60) & (0.60,\,0.60) \\
A3 & (0.45,\,0.60) & (0.55,\,0.60) & (0.70,\,0.60) \\
A4 & (0.55,\,0.60) & (0.65,\,0.60) & (0.80,\,0.60) \\
A5 & (0.65,\,0.60) & (0.75,\,0.60) & (0.85,\,0.60) \\
\end{tabular}
 \\[1em]
\midrule
\multicolumn{1}{c}{\bf Scenario 5} & \multicolumn{1}{c}{\bf Scenario 6}\\
\midrule
\begin{tabular}{lccc}
     & \textbf{B0} & \textbf{B1} & \textbf{B2} \\
\midrule
A1 & (0.30,\,0.20) & \textbf{(0.30,\,0.60)}$^{\dagger}$ & (0.45,\,0.60) \\
A2 & (0.45,\,0.30) & (0.45,\,0.60) & (0.55,\,0.60) \\
A3 & (0.55,\,0.40) & (0.55,\,0.60) & (0.66,\,0.60) \\
A4 & (0.65,\,0.60) & (0.65,\,0.60) & (0.71,\,0.60) \\
A5 & (0.75,\,0.60) & (0.75,\,0.60) & (0.75,\,0.60) \\
\end{tabular}
 & 
\begin{tabular}{lccc}
     & \textbf{B0} & \textbf{B1} & \textbf{B2} \\
\midrule
A1 & (0.15,\,0.30) & (0.20,\,0.50) & \textbf{(0.25,\,0.70)}$^{\dagger}$ \\
A2 & (0.30,\,0.30) & (0.40,\,0.50) & (0.50,\,0.70) \\
A3 & (0.45,\,0.30) & (0.50,\,0.50) & (0.60,\,0.70) \\
A4 & (0.55,\,0.30) & (0.60,\,0.50) & (0.70,\,0.70) \\
A5 & (0.65,\,0.30) & (0.70,\,0.50) & (0.80,\,0.70) \\
\end{tabular}
 \\[1em]
\midrule
\multicolumn{1}{c}{\bf Scenario 7} & \multicolumn{1}{c}{\bf Scenario 8}\\
\midrule
\begin{tabular}{lccc}
     & \textbf{B0} & \textbf{B1} & \textbf{B2} \\
\midrule
A1 & (0.01,\,0.03) & (0.03,\,0.10) & (0.05,\,0.15) \\
A2 & (0.03,\,0.05) & (0.05,\,0.20) & (0.10,\,0.25) \\
A3 & (0.05,\,0.10) & (0.10,\,0.30) & (0.15,\,0.35) \\
A4 & (0.10,\,0.20) & (0.15,\,0.40) & \textbf{(0.25,\,0.60)}$^{\dagger}$ \\
A5 & (0.15,\,0.30) & \textbf{(0.25,\,0.60)}$^{\dagger}$ & (0.45,\,0.65) \\
\end{tabular}
 & 
\begin{tabular}{lccc}
     & \textbf{B0} & \textbf{B1} & \textbf{B2} \\
\midrule
A1 & (0.01,\,0.01) & (0.03,\,0.05) & (0.05,\,0.15) \\
A2 & (0.03,\,0.05) & (0.05,\,0.15) & (0.08,\,0.25) \\
A3 & (0.05,\,0.10) & (0.08,\,0.20) & (0.10,\,0.30) \\
A4 & (0.08,\,0.15) & (0.10,\,0.25) & (0.15,\,0.40) \\
A5 & (0.10,\,0.20) & (0.15,\,0.30) & \textbf{(0.25,\,0.60)}$^{\dagger}$ \\
\end{tabular}
 \\[1em]
\midrule
\multicolumn{1}{c}{\bf Scenario 9} & \multicolumn{1}{c}{\bf Scenario 10}\\
\midrule
\begin{tabular}{lccc}
     & \textbf{B0} & \textbf{B1} & \textbf{B2} \\
\midrule
A1 & (0.05,\,0.10) & (0.10,\,0.20) & (0.15,\,0.10) \\
A2 & (0.10,\,0.20) & \textbf{(0.15,\,0.40)}$^{\dagger}$ & (0.30,\,0.20) \\
A3 & \textbf{(0.15,\,0.40)}$^{\dagger}$ & (0.30,\,0.20) & (0.45,\,0.10) \\
A4 & (0.30,\,0.20) & (0.45,\,0.10) & (0.65,\,0.05) \\
A5 & (0.45,\,0.10) & (0.65,\,0.05) & (0.75,\,0.05) \\
\end{tabular}
 & 
\begin{tabular}{lccc}
     & \textbf{B0} & \textbf{B1} & \textbf{B2} \\
\midrule
A1 & (0.10,\,0.20) & (0.80,\,0.60) & (0.90,\,0.60) \\
A2 & \textbf{(0.30,\,0.60)}$^{\dagger}$ & (0.80,\,0.60) & (0.90,\,0.60) \\
A3 & (0.55,\,0.60) & (0.80,\,0.60) & (0.90,\,0.60) \\
A4 & (0.65,\,0.60) & (0.80,\,0.60) & (0.90,\,0.60) \\
A5 & (0.75,\,0.60) & (0.80,\,0.60) & (0.90,\,0.60) \\
\end{tabular}
 \\[1em]
\midrule
\multicolumn{1}{c}{\bf Scenario 11} & \multicolumn{1}{c}{\bf Scenario 12}\\
\midrule
\begin{tabular}{lccc}
     & \textbf{B0} & \textbf{B1} & \textbf{B2} \\
\midrule
A1 & \textbf{(0.15,\,0.60)}$^{\dagger}$ & \textbf{(0.15,\,0.60)}$^{\dagger}$ & \textbf{(0.15,\,0.60)}$^{\dagger}$ \\
A2 & (0.20,\,0.60) & (0.20,\,0.60) & (0.20,\,0.60) \\
A3 & (0.25,\,0.60) & (0.25,\,0.60) & (0.25,\,0.60) \\
A4 & (0.30,\,0.60) & (0.30,\,0.60) & (0.30,\,0.60) \\
A5 & (0.35,\,0.60) & (0.35,\,0.60) & (0.35,\,0.60) \\
\end{tabular}
 & 
\begin{tabular}{lccc}
     & \textbf{B0} & \textbf{B1} & \textbf{B2} \\
\midrule
A1 & (0.01,\,0.03) & (0.03,\,0.10) & (0.05,\,0.15) \\
A2 & (0.03,\,0.05) & (0.05,\,0.20) & (0.10,\,0.25) \\
A3 & (0.05,\,0.10) & (0.10,\,0.30) & \textbf{(0.25,\,0.60)}$^{\dagger}$ \\
A4 & (0.10,\,0.20) & (0.15,\,0.40) & \textbf{(0.25,\,0.60)}$^{\dagger}$ \\
A5 & \textbf{(0.25,\,0.60)}$^{\dagger}$ & \textbf{(0.25,\,0.60)}$^{\dagger}$ & \textbf{(0.25,\,0.60)}$^{\dagger}$ \\
\end{tabular}
 \\[1em]
\bottomrule
\end{tabular}%
}
\end{table}

\clearpage
\newcommand{\beginsupplement}{%
        \setcounter{table}{0}
        \renewcommand{\thetable}{S\arabic{table}}%
        \setcounter{figure}{0}
        \renewcommand{\thefigure}{S\arabic{figure}}%
        \setcounter{section}{0}
        \renewcommand{\thesection}{S\arabic{section}}%
     }
\beginsupplement
\setcounter{page}{1}
\begin{singlespace}

\begin{center}
   \huge \textbf{Supplementary Materials} 
\end{center}

This document contains supplemental materials to the article "A seamless dose-optimization design for monotherapy and combination therapy."

\section{Bayesian optimal boundaries}\label{sup_boundaries} 
Suppose that the current cohort is treated at the dose level $j\:(j=1,...,J)$, and let $n_{j}$ denote the number of patients treated at dose level $j$. 
Let $y$ be the binary outcome. 

For toxicity, let $p_{T1} < p_{T2} < \ldots < p_{TJ}$ be the true toxicity probabilities, and let $\phi_{T0}$ denote the target toxicity probability specified by the investigator, $\phi_{T1}$ denote the highest toxicity probability that is deemed subtherapeutic such that dose-escalation should be pursued, and $\phi_{T2}$ denote the lowest toxicity probability that is deemed too toxic so that dose de-escalation is needed.
Furthermore, let $\lambda_{T1}$ and $\lambda_{T2}$ denote the lower and upper cutoffs, satisfying $\phi_{T1}<\lambda_{T1}<\phi_{T0}<\lambda_{T2}<\phi_{T2}$. 
Let $\hat{p}_{Tj}=y_{Tj}/n_{j}$ denote the observed toxicity probability based on all cumulative toxicity data at the dose level $j$, $D_{Tj}$, where $y_{Tj}$ denotes the toxicity outcomes at the dose level $j$.
We assume the following dose allocation rules for toxicity: 
\begin{itemize}
\item If $\hat{p}_{Tj}\le\lambda_{T1}$, escalate the dose level to $j+1$;
\item If $\hat{p}_{Tj}>\lambda_{T2}$, de-escalate the dose level to $j-1$;
\item If $\lambda_{T1}<\hat{p}_{Tj}\le\lambda_{T2}$, stay at the current dose $j$.
\end{itemize}
Under the Bayesian optimal boundaries for a toxicity probability, the following three hypotheses at the dose level $j$ are considered to determine the optimal values of $\lambda_{T1}$ and $\lambda_{T2}$. 
$$
H_{T0}:p_{Tj}=\phi_{T0},
H_{T1}:p_{Tj}=\phi_{T1},
H_{T2}:p_{Tj}=\phi_{T2}.
$$
Under the Bayesian paradigm, each of the hypotheses is assigned a prior probability of being true, denoted $\pi_{Ta}=\textup{Pr}(H_{Ta}), a=0,1,2$. Based on the information accumulated at dose level $j$, the probability of incorrect marginal decisions for toxicity is given by
\begin{eqnarray*}
\textup{Pr(Incorrect marginal decisions}|p_{Tj}) &=& \pi_{T0}\times \textup{Pr}(\hat{p}_{Tj}\le\lambda_{T1}~\textup{or}~\ \hat{p}_{Tj}>\lambda_{T2}|H_{T0})\\
&+& \pi_{T1}\times \textup{Pr}(\hat{p}_{Tj}>\lambda_{T1}|H_{T1})\\
&+& \pi_{T2}\times \textup{Pr}(\hat{p}_{Tj}\le\lambda_{T2}|H_{T2}).
\end{eqnarray*}
By minimizing the probability of incorrect marginal decisions for toxicity, the optimal values of $\lambda_{T1}$ and $\lambda_{T2}$ can be determined. 
A non-informative prior probability for the three hypotheses, $\pi_{T0}=\pi_{T1}=\pi_{T2}=1/3$, is specified because 1) the Bayesian optimal boundaries for toxicity are invariant to both dose level $j$ and the accumulative sample size $n_j$ and 2) the condition of $\phi_{T1}<\lambda_{T1}<\phi_{T0}<\lambda_{T2}<\phi_{T2}$ is held \citep{Liu2015-pc}.
As a result, the Bayesian optimal boundaries for toxicity are calculated as
$$
\lambda_{T1}=\textup{log}\Big(\frac{1-\phi_{T1}}{1-\phi_{T0}}\Big)\Big/\textup{log}\Big(\frac{\phi_{T0}(1-\phi_{T1})}{(1-\phi_{T0})\phi_{T1}}\Big),
$$
$$
\lambda_{T2}=\textup{log}\Big(\frac{1-\phi_{T0}}{1-\phi_{T2}}\Big)\Big/\textup{log}\Big(\frac{\phi_{T2}(1-\phi_{T0})}{(1-\phi_{T2})\phi_{T0}}\Big).
$$
In addition to the specifications of $\phi_{T0}$, the values of $\phi_{T1}$ and $\phi_{T2}$ should be specified as design parameters. The $\phi_{T1}=0.6\phi_{T0}$ and $\phi_{T2}=1.4\phi_{T0}$ (that is, 40\% deviation from the target) are recommended as the default value in the original BOIN design solely based on toxicity \citep{Liu2015-pc}.  

For efficacy, similar to \cite{Lin2017-na}, two simple hypotheses are considered to derive the optimal value for the efficacy probability. 
Let $p_{Ej}$ be the true efficacy probability at the dose level $j$, and let $\phi_{E0}$ and $\phi_{E1}$ denote the minimum required efficacy probability specified by the investigator and an inefficacious efficacy probability. 
Let $\lambda_{E}$ be the efficacy cutoff, satisfying $\phi_{E1}<\lambda_{E}<\phi_{E0}$. 
Let $\hat{p}_{Ej}=y_{Ej}/n_{j}$ denote the observed efficacy probability based on all cumulative efficacy data at the dose level $j$, $D_{Ej}$, where $y_{Ej}$ denotes the efficacy outcome at the dose level $j$.
For a tolerable dose level, the correct decisions from an efficacy perspective are described as follows. 
\begin{itemize}
\item If $\hat{p}_{Ej}\ge\lambda_{E}$, stay at the current dose level $j$; 
\item If $\hat{p}_{Ej}<\lambda_{E}$, explore other possible dose levels. 
\end{itemize}
Under the Bayesian optimal boundary for an efficacy probability, the following two hypotheses at the dose level $j$ are considered to determine the optimal value of $\lambda_{E}$. 
$$
H_{E0}:p_{Ej}=\phi_{E0},
H_{E1}:p_{Ej}=\phi_{E1},
$$
Under the Bayesian paradigm, each hypothesis is assigned a prior probability of being true, denoted as $\pi_{Ea}=\textup{Pr}(H_{Ea}),a=0,1$. Based on the information accumulated at the dose level $j$, the probability of incorrect decisions on efficacy is given by
$$
\textup{Pr(Incorrect marginal decisions}|p_{Ej})=
\pi_{E0}\times \textup{Pr}(\hat{p}_{Ej}<\lambda_{E}|H_{E0})+
\pi_{E1}\times \textup{Pr}(\hat{p}_{Ej}\ge\lambda_{E}|H_{E1}).
$$
By minimizing the probability of incorrect decisions about efficacy, the optimal value of $\lambda_{E}$ is determined. 
As with the Bayesian optimal boundaries for toxicity, a non-informative prior probability for the two hypotheses, $\pi_{E0}=\pi_{E1}=1/2$, is specified because 1) the Bayesian optimal boundary for efficacy is invariant to both dose level $j$ and the accumulative sample size $n_j$ and 2) the condition of $\phi_{E1}<\lambda_{E}<\phi_{E0}$ is held. As a result, the Bayesian optimal value of efficacy is calculated as
$$
\lambda_{E}=\textup{log}\Big(\frac{1-\phi_{E1}}{1-\phi_{E0}}\Big)\Big/\textup{log}\Big(\frac{\phi_{E0}(1-\phi_{E1})}{(1-\phi_{E0})\phi_{E1}}\Big).
$$
In addition to the specifications of $\phi_{E0}$, the value of $\phi_{E1}$ should be specified as a design parameter. The $\phi_{E1}=0.6\phi_{E0}$ (that is, the 40\% deviation from the target) is recommended as the default value of efficacy in the BOIN-ET design \citep{Takeda2018-rs}. 


\section{Likelihood with pending outcomes}\label{sup_likelihood} 
The observed toxicity data are Bernoulli variables $\tilde{y}_{Ti}, i=1,...,n_{j}$, which indicate that the patient has experienced toxicity ($\tilde{y}_{Ti}=1$), or has not yet experienced toxicity ($\tilde{y}_{Ti}=0$), at the decision times for dose-escalation or de-escalation, say $k$. 
Here, $\tilde{y}_{Ti}=1$ implies $y_{Ti}=1$, but $\tilde{y}_{Ti}=0$ implies that $\tilde{y}_{Ti}$ can be $0$ or $1$. 
Let $\gamma_{Ti}$ indicate that the toxicity outcome $y_{Ti}$ has been ascertained ($\gamma_{Ti}=1$) or is still pending ($\gamma_{Ti}=0$) by the decision time $k$, that is, when the patients in the dose-escalation cohort complete the evaluation period for toxicity, and $u_{Ti}$ ($u_{Ti}\le\tau_{T}$) denote the actual follow-up time for patient $i$ up to that moment. $\tau_{T}$ is defined as the length of the toxicity evaluation window. 
Given the interim toxicity data observed $D^{0}_{Tj}=(\tilde{y}_{T1},...,\tilde{y}_{Tn_{j}},\gamma_{T1},...,\gamma_{Tn_{j}}$) at the current dose level $j$, the joint likelihood function is given by 
\begin{eqnarray*}
L(p_{Tj}|D^{0}_{Tj}) &\propto& \prod_{i=1}^{n_{j}}p_{Tj}^{\gamma_{Ti}y_{Ti}}(1-p_{Tj})^{\gamma_{Ti}(1-y_{Ti})}(1-w_{Ti}p_{Tj})^{1-\gamma_{Ti}}\\
&=& p_{Tj}^{\tilde{x}_{Tj}}(1-p_{Tj})^{m_{Tj}}\prod_{i=1}^{n_{j}}(1-w_{Ti}p_{Tj})^{1-\gamma_{Ti}}\\
&\approx& p_{Tj}^{\tilde{x}_{Tj}}(1-p_{Tj})^{m_{Tj}}\prod_{i=1}^{n_{j}}(1-p_{Tj})^{w_{Ti}(1-\gamma_{Ti})}\\
&=& p_{Tj}^{\tilde{x}_{Tj}}(1-p_{Tj})^{\tilde{m}_{Tj}},\\
\end{eqnarray*}
where $\tilde{x}_{Tj}=\sum_{i=1}^{n_{j}}\gamma_{Ti}y_{Ti}$ is the number of patients who experienced toxicity by the interim time $k$, $\tilde{m}_{Tj}=m_{Tj}+\sum_{i=1}^{n_{j}}w_{Ti}(1-\gamma_{Ti})$ is the effective number of patients who have not experienced toxicity regardless of whether they completed the evaluation,
$m_{Tj}=\sum_{i=1}^{n_{j}}\gamma_{Ti}(1-y_{Ti})$ is the number of patients who have completed the evaluation without experiencing toxicity, and $w_{Ti}=\textup{Pr}(t_{Ti}\le u_{Ti}|y_{Ti}=1)$ can be interpreted as a weight, adjusted for the fact that the toxicity outcome has not been ascertained yet. 
The approximation in the equation for the joint likelihood function is based on the Taylor expansion of $(1-p_{Tj})^{w_{Ti}}$ at $p_{Tj}=0$ by ignoring the second-order and higher terms. 
As suggested by \citet{Lin2020-xx} and \citet{Takeda2020-yc}, we assume that the time-to-toxicity outcome is uniformly distributed over the evaluation period $(0, \tau_{T})$, that is,
$T_{Ti}|y_{Ti}=1\sim \textup{Unif}(0, \tau_{T})$, leading to $w_{Ti}=\textup{Pr}(t_{Ti}\le u_{Ti}|y_{Ti}=1)=u_{Ti}/\tau_{T}$. 
The approximate likelihood is a Bernoulli likelihood arising from the effective toxicity data $\tilde{D}_{j}=(\tilde{n}_{Tj},\tilde{x}_{Tj})$, where $\tilde{n}_{Tj}=\tilde{x}_{Tj}+\tilde{m}_{Tj}$ is the effective sample size for toxicity at the dose level $j$. 

The same approach is applied to the efficacy outcome. The observed efficacy data are quasi-Bernoulli variables $\tilde{y}_{Ei}, i=1,...,n_{j}$, which indicate that the patient has experienced efficacy ($\tilde{y}_{Ei}=1$), or has not yet experienced efficacy ($\tilde{y}_{Ei}=0$), at the decision times for dose-escalation or de-escalation, say $k$.
Here, $\tilde{y}_{Ei}=1$ implies $y_{Ei}=1$, but $\tilde{y}_{Ei}=0$ implies that $y_{Ei}$ can be $0$ or $1$. 
Let $\gamma_{Ei}$ indicate that the efficacy outcome $y_{Ei}$ has been ascertained ($\gamma_{Ei}=1$) or is still pending ($\gamma_{Ei}=0$) by the decision time $k$, and $u_{Ei}$ ($u_{Ei}\le\tau_{E}$) denote the actual follow-up time for patient $i$ up to that moment. $\tau_{E}$ is defined as the length of the efficacy evaluation window. 
Given the interim efficacy data observed $D^{0}_{Ej}=(\tilde{y}_{E1},...,\tilde{y}_{En_{j}},\gamma_{E1},...,\gamma_{En_{j}})$ at the current dose level $j$, the joint likelihood function is given by $L(p_{Ej}|D^{0}_{Ej})\propto p_{Ej}^{\tilde{x}_{Ej}}(1-p_{Ej})^{\tilde{m}_{Ej}}$, where $\tilde{x}_{Ej}=\sum_{i=1}^{n_{j}}\gamma_{Ei}y_{Ei}$ is the number of patients who experienced efficacy by the interim time $k$, $\tilde{m}_{Ej}=m_{Ej}+\sum_{i=1}^{n_{j}}w_{Ei}(1-\gamma_{Ei})$ is the effective number of patients who have not experienced efficacy regardless of whether they completed the evaluation,
$m_{Ej}=\sum_{i=1}^{n_{j}}\gamma_{Ei}(1-y_{Ei})$ is the number of patients who have completed the evaluation without experiencing efficacy and $w_{Ei}=\textup{Pr}(t_{Ei}\le u_{Ei}|y_{Ei}=1)$ can be interpreted as a weight, adjusted for the fact that the efficacy outcome has not been ascertained yet. 
The approximation in the equation for the joint likelihood function is based on the Taylor expansion of $(1-p_{Ej})^{w_{Ei}}$ at $p_{Ej}=0$ by ignoring the second-order and higher terms. 
As with the toxicity outcome, we assume that the time-to-efficacy outcome is uniformly distributed over the evaluation period $(0, \tau_{E})$, that is, 
$T_{Ei}|y_{Ei}=1\sim \textup{Unif}(0, \tau_{E})$, leading to $w_{Ei}=\textup{Pr}(t_{Ei}\le u_{Ei}|y_{Ei}=1)=u_{Ei}/\tau_{E}$. 
The approximate likelihood is a Bernoulli likelihood arising from the effective efficacy data $\tilde{D}_{Ej}=(\tilde{n}_{Ej},\tilde{x}_{Ej})$, where $\tilde{n}_{Ej}=\tilde{x}_{Ej}+\tilde{m}_{Ej}$ is the effective sample size for efficacy at the dose level $j$. 

Using effective toxicity data $\tilde{D}_{Tj}=(\tilde{n}_{Tj},\tilde{x}_{Tj})$, $\hat{p}_{Tj}$ can be replaced by $\tilde{p}_{Tj} =\tilde{x}_{Tj}/\tilde{n}_{Tj}$ as the maximum likelihood estimate of $p_{Tj}$. Using almost the same argument, with the effective efficacy data $\tilde{D}_{Ej}=(\tilde{n}_{Ej},\tilde{x}_{Ej})$, $\hat{p}_{Ej}$ can be replaced by $\tilde{p}_{Ej} =\tilde{x}_{Ej}/\tilde{n}_{Ej}$ as the maximum likelihood estimate of $p_{Ej}$.

\section{Efficacy-toxicity trade-off utility}\label{sup_utility} 
This paper applies the utility to measure the efficacy-toxicity trade-off \citep{Zhou2019-nh, Lin2020-vf} defined by
$$
U_{j}=\sum^{U}_{u=1} \psi_{u}q_{ju},
$$ 
where $\psi_{u} (u=1,...,U)$ denote utility scores attributed to the possible efficacy-toxicity outcomes and $q_{ju}$ denote the respective probabilities of observing these outcomes at dose level $j$. 
Here, $U=4$ is assumed because the efficacy and toxicity data are summarized as binary outcomes, and $u=1,2,3$, and $4$ correspond to (efficacy, no toxicity), (efficacy, toxicity), (no efficacy, no toxicity), and (no efficacy, toxicity). In this case, the most desirable outcome (no toxicity, efficacy) is assigned the highest score of $\psi_1=100$, and the least desirable outcome (toxicity, no efficacy) is assigned the lowest score of $\psi_4=0$. Given the fixed values of $\psi_1$ and $\psi_4$, the other two outcomes are assigned the scores between $\psi_1$ and $\psi_4$ to reflect clinical desirability.
The efficacy and toxicity probabilities, $p_{Ej}$ and $p_{Tj}$, are assumed to be independent, and the marginal efficacy and toxicity probabilities at dose level $j$ are $q_{j1}+q_{j2}$ and $q_{j2}+q_{j4}$, respectively.
Therefore, we have $q_{j1}=p_{Ej}(1-p_{Tj})$, $q_{j2}=p_{Ej}p_{Tj}$, $q_{j3}=(1-p_{Ej})(1-p_{Tj})$, and $q_{j4}=(1-p_{Ej})p_{Tj}$ at dose level $j$. 
During the trial, the estimate of mean utility is given by
$$
\tilde{U}_{j}=\sum^{U}_{u=1} \psi_{u}\tilde{q}_{ju},
$$ 
where $\tilde{q}_{j1}=\tilde{p}_{Ej}(1-\tilde{p}_{Tj})$, $\tilde{q}_{j2}=\tilde{p}_{Ej}\tilde{p}_{Tj}$, $\tilde{q}_{j3}=(1-\tilde{p}_{Ej})(1-\tilde{p}_{Tj})$, and $\tilde{q}_{j4}=(1-\tilde{p}_{Ej})\tilde{p}_{Tj}$ at dose level $j$ when $U=4$. 

At the end of Stage 1, all efficacy and toxicity data are used to determine the candidate doses for Stage 2. 
For toxicity, an isotonic regression is performed so that the estimated toxicity probability satisfies the monotonicity assumption. Specifically, let $\bar{p}_{Tj}$ denote the isotonic regression estimator of the observed toxicity probability at dose level $j$, $\hat{p}_{Tj}$, using the pooled adjacent violator algorithm (PAVA)\citep{Barlow-RE-and-Bartholomew-DJ-and-Bremner-JM-and-Brunk-HD1972-wh}. 
The MTD, $j_{\textup{MTD}}$, is selected as the dose whose $\bar{p}_{Tj}$ is closest to the target toxicity probability. 
For efficacy, the observed efficacy probability is used as the estimated efficacy probability such that $\bar{p}_{Ej}=\hat{p}_{Ej}$.
As a result, the estimate of mean utility when the trial is completed is given by
$$
\bar{U}_{j}=\sum^{U}_{u=1} \psi_{u}\bar{q}_{ju},
$$ 
where $\bar{q}_{j1}=\bar{p}_{Ej}(1-\bar{p}_{Tj})$, $\bar{q}_{j2}=\bar{p}_{Ej}\bar{p}_{Tj}$, $\bar{q}_{j3}=(1-\bar{p}_{Ej})(1-\bar{p}_{Tj})$, and $\bar{q}_{j4}=(1-\bar{p}_{Ej})\bar{p}_{Tj}$ at dose level $j$ when $U=4$. 
Among the $J$ dose levels that satisfy tolerability, $j \le j_{\textup{MTD}}$, the dose that maximizes the utility is selected as the OBD.

\section{Data generation using Weibull distribution}\label{sup_weibull} 
The evaluation periods for toxicity and efficacy were $\tau_{T}=28$ and $\tau_{E}=56$, as the efficacy evaluation period is typically longer than the toxicity evaluation period in oncology dose-finding studies. The accrual rate was one patient per 14 days. 
The time-to-toxicity outcomes were simulated from a Weibull distribution, $T_{i}|y_{i}=1\sim \textup{Weibull}(\zeta_{1}, \zeta_{2})$. The shape parameter $\zeta_{1}=\textup{log}\Big[\frac{\textup{log}(1-\nu_{(l,m)})}{\textup{log}(1-\nu_{(l,m)}+\alpha_{1}\nu_{(l,m)})}\Big]\Big/\textup{log}\Big(\frac{1}{1-\alpha_{2}}\Big)$ and the scale parameter $\zeta_{2}=\frac{\tau_{T}}{[-\textup{log}(1-\nu_{(l,m)})]^{1/\zeta_{1}}}$ were identified such that $\textup{Pr}(T_{i}\le\tau_{T})=\nu_{(l,m)}$ with a constraint $\textup{Pr}(T_{i}\le (1-\alpha_{2})\tau_{T})=(1-\alpha_{1})\nu_{(l,m)}$, where $\nu_{(l,m)}$ is the true probability of toxicity at the dose combination $(A_l,B_m)$, $0<\alpha_{1}<1$, and $0<\alpha_{2}<1$. As a result, the toxicity outcomes occurred with probability $\alpha_{1}$ in the last fraction of $\alpha_{2}$ of the evaluation period. In this simulation, $\alpha_{1}=0.5, \alpha_{2}=0.5$ were used; therefore, 50\% of the toxicity outcomes occurred in the second half of the evaluation period. The same approach was applied for the time-to-efficacy outcomes based on the true efficacy probability at the dose combination $(A_l,B_m)$. As a result, 50\% of the efficacy outcomes occurred in the second half of the evaluation period. To generate toxicity and efficacy outcomes, a correlation of 0.2 between toxicity and efficacy probabilities was incorporated using the Gaussian copula.

\section{Alignments with regulatory authorities}\label{sup_authority} 
The FDA Oncology Center of Excellence launched Project Optimus to reform the paradigm of dose optimization and dose selection in the development of oncology drugs and issued the guidance entitled 'Optimizing the Dosage of Human Prescription Drugs and Biological Products for the Treatment of Oncologic Diseases'~\citep{US-Food-and-Drug-Administration2024-dr}. The project aims to ensure that oncology drug doses are optimized to maximize efficacy, safety, and tolerability and to understand the pharmacokinetics, pharmacodynamics, toxicity, and efficacy at each dose level.
To achieve this goal, Project Optimus establishes the following specific objectives: (1) to communicate expectations for dose-finding and dose optimization through guidance documents, workshops, and other public meetings; (2) to provide opportunities for and encourage drug developers to meet with FDA Oncology Review Divisions early in their development programs—well before conducting trials intended for registration—to discuss dose-finding and dose optimization strategies; and (3) to develop strategies for dose-finding and dose optimization that leverage both nonclinical and clinical data in dose selection, including randomized evaluations of a range of doses in trials, with emphasis on performing these studies as early as possible in the development program and as efficiently as possible to bring promising new therapies to patients. The BOIN-MC design addresses multiple Project Optimus recommendations, including dose optimization for both monotherapy and combination therapy, and backfilling patients at lower doses to collect additional information. However, the BOIN-MC design does not currently cover: (1) integrating clinical pharmacokinetic, pharmacodynamic, and pharmacogenomic data in decision-making; (2) incorporating a randomization stratum for dose comparison; (3) considering tolerability based on less severe adverse events and patient-reported outcomes (PRO); (4) dose scheduling (i.e., the recommended interval between doses and treatment duration); and (5) identifying OBDs for multiple indications. 

In addition to Project Optimus led by the FDA, the European Medicines Agency (EMA) has also issued the guideline on the Evaluation of Anticancer Medicinal Products in Man~\citep{Committee_for_Medicinal_Products_for_Human_Use2020-gm}. This guideline addresses several EMA-specific concerns in oncology drug development, notably emphasizing the importance of integrating PK/PD data into dose selection decisions to a greater extent than the FDA guidance. Therefore, some modifications may be needed to implement the BOIN-MC design for European regulatory submissions. Additionally, requirements from other regulatory agencies, such as the Medicines and Healthcare products Regulatory Agency (MHRA) and the Pharmaceuticals and Medical Devices Agency (PMDA), as well as the revised ICH E8(R1) guideline on General Considerations for Clinical Studies, should also be considered when implementing the BOIN-MC design. Ensuring alignment with regulatory authorities remains a limitation of the BOIN-MC design implementation.

The BOIN design and its extensions are commonly used in regulatory submissions, for example, IND/IMPD. The BOIN-MC design maintains a similar level of algorithmic simplicity and incorporates comparable safety rules while demonstrating favorable operating characteristics. For practical implementation in clinical trials, investigators should include the design algorithm and key parameters in the study protocol, consistent with practices for other BOIN designs. Additionally, comprehensive simulation reports, comparable to those required for other model-assisted designs, should be submitted as part of the regulatory package.

When implementing the BOIN-MC design, investigators should ensure that participants are informed that dose allocation in the backfill cohort is partially determined by a utility function in addition to clinical judgment, in accordance with ICH E6(R3) Good Clinical Practice and the Declaration of Helsinki principles. Furthermore, the trial design, including the dose assignment process based on the utility function, should be clearly described in the protocol.

\section{Supplemental simulation results}
The eight designs were labeled as follows: (i) BOIN-MC-1, (ii) BOIN-MC-2, (iii) BOIN-W-1, (iv) BOIN-W-2, (v) BOIN-ET-1, (vi) BOIN-ET-2, (vii) BOIN12-Combo-1, and (viii) BOIN12-Combo-2.

\clearpage
\subsection{Supplemental results in simulation study}
\begin{table}[!h]
\centering
\caption{Percentage of correct OBDC selection across all part}
\label{tab:correct_obd_percent}
\begin{tabular}[t]{lrrrrrrrr}
\toprule
& \multicolumn{8}{c}{Design}\\
 & (i) & (ii) & (iii) & (iv) & (v) & (vi) & (vii) & (viii) \\
\midrule
Scenario 1 & 50.8 & 39.6 & 33.4 & 34.3 & 53.0 & 53.0 & 22.7 & 19.0
\\
Scenario 2 & 68.5 & 59.4 & 64.1 & 63.9 & 60.6 & 53.8 & 51.6 & 52.7
\\
Scenario 3 & 46.0 & 37.5 & 38.5 & 38.0 & 40.7 & 43.2 & 31.5 & 24.6
\\
Scenario 4 & 83.6 & 79.5 & 80.5 & 79.5 & 63.3 & 56.2 & 69.2 & 67.9
\\
Scenario 5 & 69.4 & 69.2 & 57.4 & 56.2 & 30.8 & 29.2 & 17.4 & 16.0
\\
Scenario 6 & 43.2 & 35.5 & 44.7 & 44.5 & 50.1 & 50.6 & 36.8 & 27.6
\\
Scenario 7 & 60.1 & 43.0 & 45.6 & 45.8 & 44.5 & 53.5 & 27.2 & 21.9
\\
Scenario 8 & 31.3 & 25.7 & 24.2 & 24.2 & 50.3 & 50.2 & 16.0 & 13.7
\\
Scenario 9 & 79.6 & 67.2 & 52.7 & 52.2 & 64.0 & 67.2 & 59.9 & 60.4
\\
Scenario 10 & 72.2 & 63.9 & 66.3 & 65.3 & 59.8 & 55.1 & 71.2 & 71.6
\\
Scenario 11 & 58.0 & 55.4 & 63.1 & 63.6 & 81.1 & 73.3 & 87.3 & 85.2
\\
Scenario 12 & 81.8 & 72.4 & 71.3 & 71.0 & 81.6 & 89.2 & 71.1 & 70.0
\\
\bottomrule
\end{tabular}
\end{table}
\begin{table}[!h]
\centering
\caption{Percentage of overdosing selection as OBDC across all part}
\label{tab:od_obd_percent}
\begin{tabular}[t]{lrrrrrrrr}
\toprule
& \multicolumn{8}{c}{Design}\\
 & (i) & (ii) & (iii) & (iv) & (v) & (vi) & (vii) & (viii) \\
\midrule
Scenario 1 & 4.3 & 2.7 & 3.5 & 3.6 & 6.6 & 11.2 & 2.2 & 2.2
\\
Scenario 2 & 17.2 & 14.9 & 11.2 & 11.8 & 25.5 & 33.1 & 15.0 & 12.2
\\
Scenario 3 & 3.5 & 3.9 & 2.6 & 2.8 & 11.6 & 12.4 & 14.6 & 14.4
\\
Scenario 4 & 5.0 & 6.6 & 5.5 & 5.8 & 14.4 & 22.0 & 9.4 & 10.8
\\
Scenario 5 & 2.6 & 2.8 & 3.1 & 3.2 & 9.7 & 11.5 & 27.6 & 28.3
\\
Scenario 6 & 4.1 & 4.4 & 4.2 & 3.8 & 9.1 & 10.3 & 9.7 & 9.9
\\
Scenario 7 & 3.0 & 2.5 & 3.2 & 3.0 & 9.7 & 8.7 & 3.1 & 2.2
\\
Scenario 8 & 0.0 & 0.0 & 0.0 & 0.0 & 0.0 & 0.0 & 0.0 & 0.0
\\
Scenario 9 & 0.2 & 0.2 & 0.1 & 0.1 & 0.2 & 0.2 & 0.4 & 0.4
\\
Scenario 10 & 1.2 & 2.0 & 2.2 & 2.2 & 13.6 & 12.3 & 3.8 & 3.4
\\
Scenario 11 & 0.8 & 0.7 & 0.6 & 0.5 & 0.4 & 0.6 & 0.0 & 0.0
\\
Scenario 12 & 0.0 & 0.0 & 0.0 & 0.0 & 0.0 & 0.0 & 0.0 & 0.0
\\
\bottomrule
\end{tabular}
\end{table}
\begin{table}[!h]
\centering
\caption{Average total number of patients across all part}
\label{tab:total_alln_rawsum}
\begin{tabular}[t]{lrrrrrrrr}
\toprule
& \multicolumn{8}{c}{Design}\\
 & (i) & (ii) & (iii) & (iv) & (v) & (vi) & (vii) & (viii) \\
\midrule
Scenario 1 & 86.0 & 63.2 & 71.8 & 71.9 & 55.3 & 54.9 & 59.7 & 59.9
\\
Scenario 2 & 65.0 & 53.2 & 63.2 & 63.6 & 54.8 & 54.9 & 57.9 & 59.5
\\
Scenario 3 & 69.5 & 57.3 & 60.1 & 59.8 & 55.0 & 54.7 & 59.1 & 59.8
\\
Scenario 4 & 49.8 & 41.6 & 39.7 & 39.6 & 39.5 & 42.0 & 52.2 & 54.5
\\
Scenario 5 & 35.6 & 31.9 & 28.2 & 27.6 & 39.0 & 39.8 & 51.7 & 53.1
\\
Scenario 6 & 57.8 & 46.7 & 45.6 & 45.7 & 46.8 & 47.4 & 56.7 & 57.4
\\
Scenario 7 & 78.2 & 63.9 & 69.4 & 69.2 & 55.7 & 55.4 & 59.8 & 60.0
\\
Scenario 8 & 74.4 & 61.7 & 67.1 & 67.2 & 55.6 & 55.2 & 60.1 & 60.1
\\
Scenario 9 & 71.4 & 57.1 & 68.1 & 68.1 & 55.7 & 55.6 & 59.2 & 59.5
\\
Scenario 10 & 35.5 & 29.0 & 18.9 & 18.7 & 24.4 & 26.9 & 33.7 & 36.4
\\
Scenario 11 & 85.8 & 57.5 & 57.6 & 57.6 & 33.6 & 34.2 & 58.5 & 58.4
\\
Scenario 12 & 75.5 & 61.1 & 69.8 & 69.8 & 55.5 & 55.1 & 60.0 & 60.1
\\
\bottomrule
\end{tabular}
\end{table}
\begin{table}[!h]
\centering
\caption{Average total trial duration (months) across all part}
\label{tab:total.duration}
\begin{tabular}[t]{lrrrrrrrr}
\toprule
& \multicolumn{8}{c}{Design}\\
 & (i) & (ii) & (iii) & (iv) & (v) & (vi) & (vii) & (viii) \\
\midrule
Scenario 1 & 25.0 & 36.0 & 40.9 & 40.9 & 55.2 & 56.1 & 54.4 & 54.5
\\
Scenario 2 & 22.2 & 30.0 & 35.6 & 35.8 & 53.5 & 54.1 & 52.4 & 53.8
\\
Scenario 3 & 23.7 & 32.6 & 34.2 & 34.0 & 54.5 & 54.6 & 52.8 & 53.3
\\
Scenario 4 & 17.0 & 23.5 & 22.5 & 22.5 & 42.3 & 45.1 & 47.0 & 49.0
\\
Scenario 5 & 14.2 & 18.2 & 16.2 & 15.9 & 39.5 & 40.4 & 41.4 & 42.4
\\
Scenario 6 & 19.4 & 26.4 & 25.8 & 25.9 & 47.6 & 48.3 & 51.4 & 51.9
\\
Scenario 7 & 25.5 & 36.9 & 40.2 & 40.1 & 53.6 & 54.4 & 54.5 & 54.6
\\
Scenario 8 & 25.0 & 36.3 & 39.4 & 39.5 & 52.6 & 52.7 & 54.4 & 54.5
\\
Scenario 9 & 23.1 & 32.4 & 38.4 & 38.4 & 52.0 & 51.9 & 53.7 & 54.1
\\
Scenario 10 & 12.2 & 16.5 & 11.5 & 11.4 & 25.6 & 28.4 & 30.1 & 32.6
\\
Scenario 11 & 23.0 & 32.9 & 33.0 & 33.0 & 36.7 & 37.3 & 53.1 & 53.1
\\
Scenario 12 & 24.7 & 35.3 & 40.3 & 40.3 & 55.2 & 56.8 & 54.5 & 54.6
\\
\bottomrule
\end{tabular}
\end{table}

\begin{table}[!h]
\centering
\caption{Percentage of OBDC selection for each dose level across all part (Scenario 1)\label{tab:obd_percent_scenario1}}
\small
\makebox[\textwidth][c]{%
%
}
\end{table}
\begin{table}[!h]
\centering
\caption{Percentage of OBDC selection for each dose level across all part (Scenario 2)\label{tab:obd_percent_scenario2}}
\small
\makebox[\textwidth][c]{%
%
}
\end{table}
\begin{table}[!h]
\centering
\caption{Percentage of OBDC selection for each dose level across all part (Scenario 3)\label{tab:obd_percent_scenario3}}
\small
\makebox[\textwidth][c]{%
%
}
\end{table}
\begin{table}[!h]
\centering
\caption{Percentage of OBDC selection for each dose level across all part (Scenario 4)\label{tab:obd_percent_scenario4}}
\small
\makebox[\textwidth][c]{%
%
}
\end{table}

\begin{table}[!h]
\centering
\caption{Percentage of OBD selection in all part (Scenario 9)\label{tab:obd_percent_scenario9}}
\small
\makebox[\textwidth][c]{%
%
}
\end{table}
\begin{table}[!h]
\centering
\caption{Percentage of OBD selection in all part (Scenario 10)\label{tab:obd_percent_scenario10}}
\small
\makebox[\textwidth][c]{%
%
}
\end{table}
\begin{table}[!h]
\centering
\caption{Percentage of OBD selection in all part (Scenario 11)\label{tab:obd_percent_scenario11}}
\small
\makebox[\textwidth][c]{%
%
}
\end{table}

\begin{table}[!h]
\centering
\caption{Results of average number of patients (Scenario 1)\label{tab:ave_alln_scenario1}}
\small
%
\end{table}
\begin{table}[!h]
\centering
\caption{Results of average number of patients (Scenario 2)\label{tab:ave_alln_scenario2}}
\small
%
\end{table}
\begin{table}[!h]
\centering
\caption{Results of average number of patients (Scenario 3)\label{tab:ave_alln_scenario3}}
\small
%
\end{table}
\begin{table}[!h]
\centering
\caption{Results of average number of patients (Scenario 4)\label{tab:ave_alln_scenario4}}
\small
%
\end{table}

\begin{table}[!h]
\centering
\caption{Percentage of correct MTD selection in mono part}
\label{tab:mono_correct_mtd_percent}
\begin{tabular}[t]{lrrrrrrrr}
\toprule
& \multicolumn{8}{c}{Design}\\
 & (i) & (ii) & (iii) & (iv) & (v) & (vi) & (vii) & (viii) \\
\midrule
Scenario 1 & 65.8 & 67.9 & 66.9 & 67.1 & 52.5 & 52.1 & 12.8 & 12.6
\\
Scenario 2 & 55.3 & 53.9 & 52.9 & 52.9 & 59.9 & 60.3 & 65.4 & 65.5
\\
Scenario 3 & 47.5 & 44.4 & 44.3 & 43.6 & 48.2 & 47.1 & 43.1 & 42.5
\\
Scenario 4 & 65.3 & 60.9 & 60.8 & 60.0 & 71.0 & 70.5 & 72.2 & 72.9
\\
Scenario 5 & 79.1 & 76.5 & 78.0 & 77.6 & 59.7 & 58.7 & 22.9 & 22.5
\\
Scenario 6 & 59.3 & 55.4 & 55.4 & 55.4 & 47.8 & 48.7 & 33.7 & 33.5
\\
Scenario 7 & 91.0 & 91.4 & 94.0 & 93.6 & 95.0 & 94.9 & 68.0 & 68.1
\\
Scenario 8 & 91.8 & 92.9 & 97.0 & 96.5 & 98.6 & 98.3 & 77.8 & 78.2
\\
Scenario 9 & 53.7 & 52.3 & 51.1 & 49.8 & 39.2 & 39.3 & 22.2 & 21.1
\\
Scenario 10 & 75.3 & 72.1 & 72.3 & 71.7 & 76.8 & 77.1 & 71.9 & 72.8
\\
Scenario 11 & 20.9 & 21.6 & 21.0 & 20.5 & 3.5 & 3.4 & 0.3 & 0.3
\\
Scenario 12 & 85.9 & 86.1 & 85.3 & 85.2 & 84.9 & 85.1 & 64.6 & 64.2
\\
\bottomrule
\end{tabular}
\end{table}

\begin{table}[!h]
\centering
\caption{Results of correct MTD selection in comb part}
\label{tab:comb_correct_mtd_percent}

\setlength{\tabcolsep}{2.5pt}
\makebox[\textwidth][c]{%
\resizebox{1.08\textwidth}{!}{%
\begin{tabular}{lrrrrrrrr}
\toprule
& \multicolumn{8}{c}{Design}\\
 & (i) & (ii) & (iii) & (iv) & (v) & (vi) & (vii) & (viii) \\
\midrule
Scenario 1 & (66.7, 56.4) & (65.4, 55.2) & (57.9, 56.8) & (59.9, 57.4) & (47.8, 34.0) & (46.2, 42.9) & (7.8, 38.2) & (16.4, 28.5) \\
Scenario 2 & (31.8, 26.7) & (34.3, 29.2) & (56.4, 55.5) & (56.6, 54.9) & (61.5, 65.6) & (61.3, 56.2) & (15.4, 22.6) & (17.8, 16.2) \\
Scenario 3 & (43.8, 61.0) & (45.0, 58.6) & (48.0, 58.1) & (47.7, 57.6) & (40.4, 44.2) & (40.4, 46.1) & (18.3, 44.5) & (28.0, 34.6) \\
Scenario 4 & (81.8, 42.7) & (76.4, 42.1) & (71.9, 13.5) & (71.0, 13.8) & (79.2, 67.7) & (63.0, 64.2) & (36.7, 13.3) & (25.0, 11.8) \\
Scenario 5 & (97.4, 23.4) & (95.4, 24.5) & (59.7, 16.6) & (58.5, 16.7) & (49.5, 29.9) & (47.8, 29.9) & (29.5, 12.5) & (23.8, 11.8) \\
Scenario 6 & (68.3, 66.4) & (61.8, 63.9) & (38.8, 58.7) & (39.2, 57.8) & (68.4, 71.4) & (63.3, 71.4) & (15.4, 47.1) & (13.5, 43.1) \\
Scenario 7 & (68.5, 66.2) & (71.8, 62.3) & (81.2, 53.6) & (80.8, 53.7) & (60.6, 25.3) & (60.6, 47.5) & (11.2, 52.6) & (27.9, 40.9) \\
Scenario 8 & (85.0, 69.5) & (86.2, 70.7) & (95.0, 67.8) & (94.8, 67.4) & (91.4, 82.4) & (90.4, 77.2) & (10.4, 43.1) & (12.0, 46.9) \\
Scenario 9 & (64.9, 49.8) & (61.3, 49.8) & (57.8, 54.3) & (57.5, 54.4) & (47.2, 12.6) & (47.8, 18.9) & (20.8, 33.2) & (20.8, 32.3) \\
Scenario 10 & (0.0, 0.0) & (0.0, 0.0) & (0.0, 0.0) & (0.0, 0.0) & (0.0, 0.0) & (0.0, 0.0) & (0.0, 0.0) & (0.0, 0.0) \\
Scenario 11 & (10.3, 6.1) & (14.8, 8.9) & (16.4, 8.9) & (16.7, 9.0) & (0.2, 0.0) & (1.8, 0.8) & (0.0, 0.0) & (0.1, 0.1) \\
Scenario 12 & (63.5, 37.9) & (66.6, 39.8) & (73.0, 28.7) & (70.8, 28.1) & (54.2, 22.7) & (64.3, 24.0) & (6.8, 9.0) & (26.0, 18.4) \\
\bottomrule
\end{tabular}%
}}
\end{table}

\begin{table}[!h]
\centering
\caption{Percentage of correct OBD selection in comb part}
\label{tab:comb_correct_obd_percent}
\begin{tabular}[t]{lrrrrrrrr}
\toprule
& \multicolumn{8}{c}{Design}\\
 & (i) & (ii) & (iii) & (iv) & (v) & (vi) & (vii) & (viii) \\
\midrule
Scenario 1 & 62.0 & 54.3 & 39.5 & 40.6 & 42.9 & 47.1 & 38.7 & 35.4
\\
Scenario 2 & 40.5 & 44.0 & 65.4 & 65.3 & 63.7 & 52.5 & 43.9 & 52.8
\\
Scenario 3 & 48.7 & 43.0 & 41.7 & 41.1 & 44.3 & 47.3 & 60.0 & 58.2
\\
Scenario 4 & 79.2 & 78.7 & 92.0 & 91.7 & 75.3 & 59.8 & 47.0 & 40.8
\\
Scenario 5 & 79.9 & 80.3 & 72.8 & 71.9 & 88.0 & 84.3 & 38.1 & 33.5
\\
Scenario 6 & 46.0 & 40.2 & 48.9 & 48.6 & 52.2 & 52.9 & 50.8 & 45.3
\\
Scenario 7 & 62.5 & 50.0 & 51.2 & 51.7 & 46.8 & 56.3 & 58.1 & 54.2
\\
Scenario 8 & 32.5 & 29.0 & 26.1 & 26.1 & 53.0 & 52.8 & 32.2 & 29.4
\\
Scenario 9 & 53.8 & 55.1 & 33.7 & 32.4 & 60.1 & 61.6 & 54.1 & 20.4
\\
Scenario 10 & 99.3 & 99.6 & 99.8 & 99.7 & 99.9 & 99.2 & 0.0 & 0.0
\\
Scenario 11 & 57.0 & 54.8 & 70.7 & 70.8 & 90.4 & 78.6 & 87.0 & 83.7
\\
Scenario 12 & 76.0 & 70.1 & 69.3 & 67.2 & 65.8 & 85.6 & 77.8 & 77.4
\\
\bottomrule
\end{tabular}
\end{table}


\clearpage
\subsection{Standard deviation of continuous outcome for simulation study}
\begin{table}[!h]
\centering
\caption{Standard deviation of the total trial duration (months) across all part}
\label{tab:sd_total_duration_all_part}
\begin{tabular}[t]{lrrrrrrrr}
\toprule
& \multicolumn{8}{c}{Design}\\
 & (i) & (ii) & (iii) & (iv) & (v) & (vi) & (vii) & (viii) \\
\midrule
Scenario 1 & 4.4 & 4.7 & 3.9 & 3.9 & 4.9 & 5.4 & 2.5 & 1.1 \\
Scenario 2 & 6.7 & 6.9 & 6.6 & 6.6 & 7.3 & 7.5 & 7.3 & 4.8 \\
Scenario 3 & 3.6 & 3.9 & 6.7 & 6.8 & 13.6 & 13.7 & 8.7 & 8.0 \\
Scenario 4 & 3.8 & 4.3 & 6.0 & 6.0 & 12.1 & 12.6 & 12.0 & 10.8 \\
Scenario 5 & 3.6 & 4.0 & 7.0 & 7.0 & 15.8 & 16.4 & 18.0 & 17.8 \\
Scenario 6 & 4.3 & 4.7 & 7.1 & 7.0 & 14.0 & 14.0 & 9.6 & 9.0 \\
Scenario 7 & 3.4 & 3.7 & 2.8 & 2.9 & 2.9 & 3.2 & 1.7 & 1.3 \\
Scenario 8 & 3.4 & 3.5 & 2.2 & 2.2 & 2.6 & 2.5 & 1.5 & 0.9 \\
Scenario 9 & 4.2 & 4.8 & 5.5 & 5.6 & 6.3 & 6.1 & 4.8 & 3.5 \\
Scenario 10 & 0.6 & 0.6 & 2.8 & 2.8 & 6.4 & 6.9 & 3.5 & 4.3 \\
Scenario 11 & 4.9 & 5.2 & 9.2 & 9.3 & 11.7 & 12.4 & 8.0 & 7.5 \\
Scenario 12 & 2.9 & 3.1 & 2.9 & 3.0 & 2.8 & 3.6 & 1.6 & 1.4 \\
\bottomrule
\end{tabular}
\end{table}

\begin{table}[!h]
\centering
\caption{Standard deviation of the total number of patients across all part}
\label{tab:sd_total_n_all_part}
\begin{tabular}[t]{lrrrrrrrr}
\toprule
& \multicolumn{8}{c}{Design}\\
 & (i) & (ii) & (iii) & (iv) & (v) & (vi) & (vii) & (viii) \\
\midrule
Scenario 1 & 18.6 & 11.9 & 7.5 & 7.4 & 4.3 & 4.7 & 2.7 & 1.2 \\
Scenario 2 & 20.1 & 16.2 & 12.7 & 12.8 & 7.2 & 7.4 & 8.0 & 5.2 \\
Scenario 3 & 18.7 & 12.9 & 12.9 & 13.0 & 13.0 & 13.0 & 9.5 & 8.7 \\
Scenario 4 & 15.9 & 12.1 & 11.7 & 11.8 & 10.9 & 11.2 & 13.0 & 11.7 \\
Scenario 5 & 15.0 & 12.6 & 13.5 & 13.7 & 14.8 & 15.4 & 19.6 & 19.4 \\
Scenario 6 & 20.4 & 15.0 & 13.8 & 13.7 & 13.3 & 13.4 & 10.4 & 9.7 \\
Scenario 7 & 16.1 & 9.6 & 5.1 & 5.1 & 2.8 & 3.0 & 1.9 & 1.5 \\
Scenario 8 & 15.5 & 9.1 & 3.9 & 3.9 & 2.7 & 2.6 & 1.6 & 1.0 \\
Scenario 9 & 16.0 & 12.4 & 10.6 & 10.8 & 6.7 & 6.6 & 5.2 & 3.8 \\
Scenario 10 & 8.9 & 5.3 & 5.4 & 5.4 & 5.8 & 6.2 & 3.8 & 4.7 \\
Scenario 11 & 32.4 & 17.4 & 17.2 & 17.3 & 10.6 & 11.2 & 8.6 & 8.1 \\
Scenario 12 & 13.8 & 8.0 & 5.3 & 5.4 & 2.8 & 3.5 & 1.7 & 1.5 \\
\bottomrule
\end{tabular}
\end{table}

\clearpage
\subsection{Different utility scores}
\begin{table}[!htbp]
\centering
\caption{Comparison of correct OBD selection percentages}
\label{tab:correct-obd-percent-compare}
\begin{tabular}{ccccc}
\toprule
 & \multicolumn{2}{c}{Original utility scores} & \multicolumn{2}{c}{Different utility scores} \\
 & \multicolumn{2}{c}{(100,60,40,0)} & \multicolumn{2}{c}{(100,75,25,0)} \\
\cmidrule(lr){2-3} \cmidrule(lr){4-5}
Scenario & BOIN-MC1 & BOIN-MC2 & BOIN-MC1 & BOIN-MC2 \\
\midrule
1 & 50.8 & 39.6 & 60.0 & 46.0 \\
2 & 68.5 & 59.4 & 63.8 & 58.3 \\
3 & 46.0 & 37.5 & 51.0 & 42.4 \\
4 & 83.6 & 79.5 & 80.8 & 76.7 \\
5 & 69.4 & 69.2 & 71.3 & 71.1 \\
6 & 43.2 & 35.5 & 47.8 & 39.7 \\
7 & 60.1 & 43.0 & 65.9 & 49.7 \\
8 & 31.3 & 25.7 & 39.9 & 32.6 \\
9 & 79.6 & 67.2 & 80.9 & 69.6 \\
10 & 72.2 & 63.9 & 69.4 & 62.9 \\
11 & 58.0 & 55.4 & 52.8 & 52.8 \\
12 & 81.8 & 72.4 & 87.3 & 78.6 \\
\bottomrule
\end{tabular}
\end{table}

\subsection{Different correlations}
\begin{table}[!htbp]
\centering
\caption{Comparison of correct OBD selection percentages in difference correlations}
\label{tab:correct-obd-percent-compare-correlation}
\begin{tabular}{ccccccc}
\toprule
 & \multicolumn{2}{c}{Original} & \multicolumn{2}{c}{Different 1} & \multicolumn{2}{c}{Different 2} \\
 & \multicolumn{2}{c}{Correlation of 0.2} & \multicolumn{2}{c}{Correlation of 0} & \multicolumn{2}{c}{Correlation of -0.2} \\
\cmidrule(lr){2-3} \cmidrule(lr){4-5} \cmidrule(lr){6-7}
Scenario & BOIN-MC1 & BOIN-MC2 & BOIN-MC1 & BOIN-MC2 & BOIN-MC1 & BOIN-MC2 \\
\midrule
1 & 50.8 & 39.6 & 50.5 & 37.8 & 50.9 & 36.7 \\
2 & 68.5 & 59.4 & 68.4 & 59.5 & 67.4 & 58.7 \\
3 & 46.0 & 37.5 & 46.2 & 36.2 & 44.8 & 36.0 \\
4 & 83.6 & 79.5 & 83.3 & 78.6 & 81.8 & 78.4 \\
5 & 69.4 & 69.2 & 69.4 & 68.7 & 67.7 & 67.0 \\
6 & 43.2 & 35.5 & 41.3 & 33.5 & 40.2 & 32.6 \\
7 & 60.1 & 43.0 & 59.4 & 42.6 & 57.5 & 41.7 \\
8 & 31.3 & 25.7 & 31.0 & 23.5 & 30.8 & 22.5 \\
9 & 79.6 & 67.2 & 78.9 & 67.0 & 78.7 & 65.2 \\
10 & 72.2 & 63.9 & 73.0 & 63.8 & 70.4 & 63.8 \\
11 & 58.0 & 55.4 & 57.3 & 55.3 & 57.9 & 55.5 \\
12 & 81.8 & 72.4 & 81.7 & 71.4 & 80.8 & 70.8 \\
\bottomrule
\end{tabular}
\end{table}

\subsection{Different starting dose-combination}
\begin{table}[!htbp]
\centering
\caption{Comparison of correct OBD selection percentages under two starting dose rules for combination subtrials}
\label{tab:correct-obd-percent-compare-dose}
\begin{tabular}{ccccc}
\toprule
 & \multicolumn{2}{c}{Starting at (OBD-1,1)} & \multicolumn{2}{c}{Starting at (1,1)} \\
\cmidrule(lr){2-3} \cmidrule(lr){4-5}
Scenario & BOIN-MC1 & BOIN-MC2 & BOIN-MC1 & BOIN-MC2 \\
\midrule
1 & 50.8 & 39.6 & 43.7 & 27.0 \\
2 & 68.5 & 59.4 & 74.2 & 64.8 \\
3 & 46.0 & 37.5 & 39.6 & 31.0 \\
4 & 83.6 & 79.5 & 83.8 & 80.0 \\
5 & 69.4 & 69.2 & 70.6 & 69.6 \\
6 & 43.2 & 35.5 & 45.2 & 36.5 \\
7 & 60.1 & 43.0 & 50.3 & 32.8 \\
8 & 31.3 & 25.7 & 29.5 & 21.5 \\
9 & 79.6 & 67.2 & 73.3 & 61.7 \\
10 & 72.2 & 63.9 & 72.1 & 63.9 \\
11 & 58.0 & 55.4 & 65.7 & 64.2 \\
12 & 81.8 & 72.4 & 79.5 & 60.2 \\
\bottomrule
\end{tabular}
\end{table}

\clearpage
\subsection{Probabilities of edge-case stopping events}
\begin{table}[h]
\centering
\caption{Probabilities of edge-case stopping events}
\label{tab:edge-case-stop-probability}
\begin{tabular}{c ccc ccc}
\toprule
 & \multicolumn{3}{c}{BOIN-MC1} & \multicolumn{3}{c}{BOIN-MC2} \\
\cmidrule(lr){2-4} \cmidrule(lr){5-7}
 & (a) & (b) & (c) & (a) & (b) & (c) \\
\midrule
Scenario 1 & 0.3 & 15.2 & 4.9 & - & 14.7 & 3.6 \\
Scenario 2 & 0.1 & 45.4 & 6.4 & - & 43.5 & 6.2 \\
Scenario 3 & 0.2 & 5.4 & 8.9 & - & 6.9 & 8.0 \\
Scenario 4 & 0.0 & 25.1 & 1.2 & - & 27.9 & 1.1 \\
Scenario 5 & 0.0 & 12.9 & 0.9 & - & 14.3 & 0.6 \\
Scenario 6 & 0.0 & 18.2 & 3.7 & - & 20.8 & 3.6 \\
Scenario 7 & 1.5 & 4.2 & 5.3 & - & 4.2 & 4.2 \\
Scenario 8 & 2.0 & 0.8 & 30.3 & - & 0.5 & 25.4 \\
Scenario 9 & 2.4 & 13.7 & 39.9 & - & 11.9 & 33.1 \\
Scenario 10 & 0.0 & 99.4 & 0.0 & - & 99.2 & 0.0 \\
Scenario 11 & 0.0 & 4.7 & 1.2 & - & 5.6 & 0.9 \\
Scenario 12 & 0.2 & 4.9 & 5.8 & - & 4.6 & 5.2 \\
\bottomrule
\end{tabular}
\begin{flushleft}
\footnotesize
For BOIN-MC2, reason (a) is shown as "-" because no backfill cohort is used.
\end{flushleft}
\end{table}

\subsection{Additional Safety Operating Characteristics}
\begin{table}[!h]
\centering
\caption{Average number of patients treated above the true MTDC across all parts}
\label{tab:n_above_true_mtdc}
\begin{tabular}[t]{lrrrrrrrr}
\toprule
& \multicolumn{8}{c}{Design}\\
 & (i) & (ii) & (iii) & (iv) & (v) & (vi) & (vii) & (viii) \\
\midrule
Scenario 1 & 7.4 & 7.9 & 8.0 & 8.1 & 1.7 & 4.8 & 1.0 & 2.3
\\
Scenario 2 & 19.2 & 18.7 & 17.7 & 18.1 & 12.3 & 16.7 & 10.7 & 16.9
\\
Scenario 3 & 8.8 & 9.0 & 8.7 & 8.6 & 9.9 & 10.6 & 6.8 & 8.1
\\
Scenario 4 & 10.1 & 10.6 & 8.5 & 8.7 & 6.6 & 10.4 & 4.6 & 8.7
\\
Scenario 5 & 7.5 & 7.6 & 7.2 & 7.2 & 11.8 & 12.7 & 11.4 & 13.2
\\
Scenario 6 & 14.1 & 15.1 & 13.7 & 13.6 & 13.6 & 14.9 & 7.8 & 9.6
\\
Scenario 7 & 5.3 & 5.3 & 3.9 & 3.9 & 1.7 & 2.1 & 1.5 & 2.4
\\
Scenario 8 & 0.0 & 0.0 & 0.0 & 0.0 & 0.0 & 0.0 & 0.0 & 0.0
\\
Scenario 9 & 9.6 & 10.5 & 12.2 & 12.1 & 8.4 & 9.3 & 4.5 & 4.7
\\
Scenario 10 & 4.7 & 4.8 & 4.5 & 4.6 & 3.5 & 3.5 & 1.5 & 1.4
\\
Scenario 11 & 3.2 & 3.5 & 3.8 & 3.9 & 0.2 & 0.3 & 0.0 & 0.0
\\
Scenario 12 & 0.0 & 0.0 & 0.0 & 0.0 & 0.0 & 0.0 & 0.0 & 0.0
\\
\bottomrule
\end{tabular}
\end{table}

\begin{table}[!h]
\centering
\caption{Percentage of trials terminated early for toxicity across all part}
\label{tab:earlystop_tox_percent}
\begin{tabular}[t]{lrrrrrrrr}
\toprule
& \multicolumn{8}{c}{Design}\\
 & (i) & (ii) & (iii) & (iv) & (v) & (vi) & (vii) & (viii) \\
\midrule
Scenario 1 & 17.7 & 16.5 & 0.1 & 0.0 & 3.0 & 9.0 & 0.8 & 0.4
\\
Scenario 2 & 49.6 & 47.4 & 6.8 & 7.0 & 9.5 & 13.6 & 6.7 & 4.5
\\
Scenario 3 & 7.1 & 8.0 & 0.9 & 1.3 & 6.6 & 7.6 & 2.7 & 1.5
\\
Scenario 4 & 26.8 & 28.7 & 24.4 & 24.2 & 48.1 & 46.6 & 33.5 & 31.6
\\
Scenario 5 & 26.7 & 27.9 & 27.1 & 27.8 & 24.3 & 23.8 & 26.1 & 25.0
\\
Scenario 6 & 19.8 & 22.5 & 21.7 & 22.4 & 8.9 & 8.8 & 12.1 & 10.8
\\
Scenario 7 & 5.7 & 5.4 & 0.0 & 0.0 & 1.0 & 3.7 & 0.3 & 0.1
\\
Scenario 8 & 1.6 & 1.1 & 0.0 & 0.0 & 0.2 & 0.5 & 0.2 & 0.1
\\
Scenario 9 & 24.9 & 23.0 & 0.8 & 1.0 & 2.5 & 4.5 & 2.7 & 2.6
\\
Scenario 10 & 99.5 & 99.4 & 98.9 & 98.9 & 79.4 & 79.4 & 95.4 & 95.4
\\
Scenario 11 & 6.1 & 6.7 & 3.7 & 3.4 & 0.9 & 0.9 & 6.4 & 5.9
\\
Scenario 12 & 6.8 & 6.0 & 0.0 & 0.0 & 1.9 & 7.4 & 0.3 & 0.0
\\
\bottomrule
\end{tabular}
\end{table}

\subsection{Average number of concurrently active dose levels}
\begin{table}[h]
\centering
\caption{Average number of concurrently active dose levels.}
\label{tab:ave_concurrent_dose}
\begin{tabular}{|c|c|c|}
\hline
Scenario & BOIN-MC1 & BOIN-MC2 \\
\hline
1 & 1.56 & 1.32 \\
2 & 1.46 & 1.32 \\
3 & 1.43 & 1.30 \\
4 & 1.43 & 1.30 \\
5 & 1.28 & 1.21 \\
6 & 1.44 & 1.28 \\
7 & 1.47 & 1.30 \\
8 & 1.44 & 1.29 \\
9 & 1.53 & 1.36 \\
10 & 1.41 & 1.30 \\
11 & 1.54 & 1.24 \\
12 & 1.45 & 1.29 \\
\hline
\end{tabular}
\end{table}

\clearpage
\subsection{Different matrix size and results}
\begin{table}[h]
\centering
\caption{True toxicity and efficacy probabilities. The probabilities at OBDCs across both monotherapy and combination therapy were in boldface. The red asterisk marks the dose level considered to be the OBDC based on monotherapy alone in cases where no OBDC was identified for monotherapy when considering the overall results.}
\label{tab:tox_eff_scenarios_diff}
\resizebox{\textwidth}{!}{%
\begin{tabular}{cc}
\toprule
\multicolumn{1}{c}{\textbf{Scenario 13}} & \multicolumn{1}{c}{\textbf{Scenario 14}}\\
\midrule
\begin{tabular}{lccc}
     & \textbf{B0} & \textbf{B1} & \textbf{B2} \\
\midrule
A1 & $(0.04,\,0.20)$ & $(0.10,\,0.30)$ & $(0.13,\,0.40)$ \\
A2 & $(0.11,\,0.30)$ & $(0.15,\,0.40)$ & $(\mathbf{0.30},\,\mathbf{0.60})$ \\
A3 & $(0.15,\,0.40)$ & $(\mathbf{0.30},\,\mathbf{0.60})$ & $(0.45,\,0.65)$ \\
A4 & $(\mathbf{0.30},\,\mathbf{0.60})$ & $(0.45,\,0.65)$ & $(0.55,\,0.70)$ \\
\end{tabular}
 & 
\begin{tabular}{lccc}
     & \textbf{B0} & \textbf{B1} & \textbf{B2} \\
\midrule
A1 & $(0.10,\,0.10)$ & $(\mathbf{0.25},\,\mathbf{0.45})$ & $(0.45,\,0.50)$ \\
A2 & $(0.15,\,0.20)$ & $(0.45,\,0.50)$ & $(0.55,\,0.55)$ \\
A3 & $(\mathbf{0.25},\,\mathbf{0.45})$ & $(0.55,\,0.55)$ & $(0.65,\,0.60)$ \\
A4 & $(0.45,\,0.50)$ & $(0.65,\,0.60)$ & $(0.75,\,0.65)$ \\
\end{tabular}
 \\[1em]
\midrule
\multicolumn{1}{c}{\textbf{Scenario 15}} & \multicolumn{1}{c}{\textbf{Scenario 16}}\\
\midrule
\begin{tabular}{lcccc}
     & \textbf{B0} & \textbf{B1} & \textbf{B2} & \textbf{B3} \\
\midrule
A1 & $(0.01,\,0.10)$ & $(0.05,\,0.20)$ & $(0.07,\,0.30)$ & $(0.10,\,0.40)$ \\
A2 & $(0.04,\,0.20)$ & $(0.10,\,0.30)$ & $(0.13,\,0.40)$ & $(\mathbf{0.30},\,\mathbf{0.60})$ \\
A3 & $(0.11,\,0.30)$ & $(0.15,\,0.40)$ & $(\mathbf{0.30},\,\mathbf{0.60})$ & $(0.45,\,0.65)$ \\
A4 & $(0.15,\,0.40)$ & $(\mathbf{0.30},\,\mathbf{0.60})$ & $(0.45,\,0.65)$ & $(0.55,\,0.70)$ \\
A5 & $(\mathbf{0.30},\,\mathbf{0.60})$ & $(0.45,\,0.65)$ & $(0.55,\,0.70)$ & $(0.65,\,0.75)$ \\
\end{tabular}
 & 
\begin{tabular}{lcccc}
     & \textbf{B0} & \textbf{B1} & \textbf{B2} & \textbf{B3} \\
\midrule
A1 & $(0.05,\,0.05)$ & $(0.15,\,0.15)$ & $(\mathbf{0.25},\,\mathbf{0.45})$ & $(0.45,\,0.50)$ \\
A2 & $(0.10,\,0.10)$ & $(\mathbf{0.25},\,\mathbf{0.45})$ & $(0.45,\,0.50)$ & $(0.55,\,0.55)$ \\
A3 & $(0.15,\,0.20)$ & $(0.45,\,0.50)$ & $(0.55,\,0.55)$ & $(0.65,\,0.60)$  \\
A4 & $(\mathbf{0.25},\,\mathbf{0.45})$ & $(0.55,\,0.55)$ & $(0.65,\,0.60)$ & $(0.75,\,0.65)$ \\
A5 & $(0.45,\,0.50)$ & $(0.65,\,0.60)$ & $(0.75,\,0.65)$ & $(0.80,\,0.70)$ \\
\end{tabular}
 \\
\bottomrule
\end{tabular}
}
\end{table}

\begin{table}[!h]
\centering
\caption{Percentage of correct OBD selection in all part}
\label{tab:correct_obd_percent_diff_mat}
\begin{tabular}[t]{lrrrrrrrr}
\toprule
& \multicolumn{8}{c}{Design}\\
 & (i) & (ii) & (iii) & (iv) & (v) & (vi) & (vii) & (viii) \\
\midrule
Scenario 13 & 52.1 & 39.7 & 36.5 & 37.5 & 46.1 & 51.8 & 26.1 & 22.4
\\
Scenario 14 & 67.6 & 59.5 & 64.0 & 62.9 & 34.7 & 26.0 & 55.3 & 55.6
\\
Scenario 15 & 49.5 & 39.0 & 38.4 & 38.3 & 49.8 & 54.7 & 28.7 & 24.6
\\
Scenario 16 & 65.5 & 57.2 & 60.2 & 59.4 & 41.6 & 38.0 & 49.6 & 49.7
\\
\bottomrule
\end{tabular}
\end{table}

\clearpage
\subsection{Operating characteristics for the small sample sizes (n=3, 6)}
\begin{table}[htbp]
\centering
\caption{Probabilities of dose-escalation, staying at the current dose,
and dose de-escalation based on the toxicity boundary when three
patients have been evaluated at the current dose level.}
\label{tab:toxicity-decision-probability-n3}
\begin{threeparttable}
\begin{tabular}{rrrr}
\toprule
True toxicity probability
& Escalate
& Stay
& De-escalate \\
\midrule
0.30 & 34.3\% & 44.1\% & 21.6\% \\
0.40 & 21.6\% & 43.2\% & 35.2\% \\
0.50 & 12.5\% & 37.5\% & 50.0\% \\
0.60 &  6.4\% & 28.8\% & 64.8\% \\
0.70 &  2.7\% & 18.9\% & 78.4\% \\
0.80 &  0.8\% &  9.6\% & 89.6\% \\
0.90 &  0.1\% &  2.7\% & 97.2\% \\
\bottomrule
\end{tabular}
\end{threeparttable}
\end{table}

\begin{table}[htbp]
\centering
\caption{Probabilities of dose-escalation, staying at the current dose,
and dose de-escalation based on the toxicity boundary when six
patients have been evaluated at the current dose level.}
\label{tab:toxicity-decision-probability-n6}
\begin{threeparttable}
\begin{tabular}{rrrr}
\toprule
True toxicity probability
& Escalate
& Stay
& De-escalate \\
\midrule
0.30 & 42.02\% & 32.41\% & 25.57\% \\
0.40 & 23.33\% & 31.10\% & 45.57\% \\
0.50 & 10.94\% & 23.44\% & 65.63\% \\
0.60 &  4.10\% & 13.82\% & 82.08\% \\
0.70 &  1.09\% &  5.95\% & 92.95\% \\
0.80 &  0.16\% &  1.54\% & 98.30\% \\
0.90 &  0.01\% &  0.12\% & 99.87\% \\
\bottomrule
\end{tabular}
\end{threeparttable}
\end{table}

\end{singlespace}

\end{document}